\documentclass{article}\begin{document}\centerline{\bf Schrodinger's Equation in Riemann Spaces\rm}
\[
\]
\centerline{\bf Nikos Bagis}

\centerline{Department of Informatics}

\centerline{Aristotle University of Thessaloniki Greece}
\centerline{nikosbagis@hotmail.gr}

\begin{quote}

\begin{abstract}

We present some properties of the first and second order Beltrami differential operators in Riemann metric spaces. We also solve the Schrodinger's equation for a wide class of potentials and describe when the Hamiltonian of a physical system is self adjoint.    

\end{abstract}

\bf keywords \rm{Schrodinger's Equation; Metric Space; Riemann Geometry; Differential Operators; Laplacian}

\end{quote}

\section{Introduction}

In a $3-$dimensional space, in the case of orthogonal coordinate system $(x,y,z)$ of elementary geometry we make use of the differential operator:   
\begin{equation}
\Delta_2\Phi=\frac{\partial^2\Phi}{\partial x^2}+\frac{\partial^2\Phi}{\partial y^2}+\frac{\partial^2\Phi}{\partial z^2}.
\end{equation}
We call this differential operator, Laplacian operator, or Laplace-Beltrami differential parameter of second kind. One of many uses of this operator, is that help us describe the state of a particle in $\bf E_3\rm$, (the Eucledian three dimensional space). If in $\bf E_3\rm$ exist also a potential $V=V(x,y,z,t)$, depending on position and time, then the Schroedinger's equation of the particle read as
\begin{equation}
i \hbar \frac{\partial\Psi}{\partial t}=-\frac{\hbar^2}{2m}\left(\frac{\partial^2\Psi}{\partial x^2}+\frac{\partial^2\Psi}{\partial y^2}+\frac{\partial^2\Psi}{\partial z^2}\right)+V\Psi
\end{equation}  
and $\Psi=\Psi(x,y,z,t)$ describes the state of the system.\\

In the present article we consider the case of a $N$ dimensional metric space $S$ of curvilinear coordinates $x_l$
$$
\overline{\bf S\rm}(x_1,x_2,\ldots,x_N)=\{S_1(x_1,x_2,\ldots x_N),S_2(x_1,x_2,\ldots ,x_N),\dots,
$$
\begin{equation}
S_N(x_1,x_2,\ldots,x_N)\}
\end{equation}
with metric tensor 
\begin{equation}
g_{ik}=g_{ik}(x^a)=\left\langle \frac{\partial \overline{\bf  S\rm}}{\partial x_i},\frac{\partial \overline{\bf S\rm}}{\partial x_k}\right\rangle_{reg}.
\end{equation}
Note that the above product is the regular dot product of the vectors 
$$
\frac{\partial\overline{\bf S\rm}}{\partial x_i}=\left\{\frac{\partial S_1(x^a)}{\partial x_i},\frac{\partial S_2(x^a)}{\partial x_i},\ldots,\frac{\partial S_N(x^a)}{\partial x_i}\right\}
$$ 
and 
$$
\frac{\partial\overline{\bf S\rm}}{\partial x_k}=\left\{\frac{\partial S_1(x^a)}{\partial x_k},\frac{\partial S_2(x^a)}{\partial x_k},\ldots,\frac{\partial S_N(x^a)}{\partial x_k}\right\},
$$ 
i.e. for $\overline{A}=\{A_1,A_2,\ldots,A_N\}$ and $\overline{B}=\{B_1,B_2,\ldots,B_N\}$ we have 
$$
\left\langle \overline{A},\overline{B}\right\rangle_{reg}=\sum^N_{i=1}A_iB_i.
$$
Also $g^{ij}=(g_{ij})^t=(g_{ij})^{(-1)}$ and obviously $g_{ij}=g_{ji}$ and $x^a$ denotes the vector  $\{x_1,x_2,\ldots,x_N\}$.\\The linear element will be
\begin{equation}
ds^2=\sum^{N}_{i,j=1}g_{ij}dx_idx_j ,
\end{equation}
where $g^{ij}=(g_{ij})^t=(g_{ij})^{(-1)}$ and $g_{ij}=g_{ji}$. With the help of the above metric we define\\
\\  
\textbf{Definition 1.}\\ The 1-Beltrami differential operator is 
\begin{equation}
\Delta_1(\Phi,\Psi):=\sum^{N}_{i,j=1}(g_{ij})^t\left(\partial_i\Phi \partial_j\Psi+\partial_j\Phi \partial_i\Psi\right)=2\sum^{N}_{i,j=1}(g_{ij})^t\partial_i\Phi \partial_j\Psi.
\end{equation}
\\
\textbf{Definition 2.}\\The 2-Beltrami operator is
\begin{equation}
\Delta_2\Phi=\sum^{N}_{i,j=1}(g_{ij})^t\left(\frac{\partial^2\Phi}{\partial u_i\partial u_j}-\sum^{N}_{k=1}\Gamma^k_{ij}\partial_k\Phi\right),
\end{equation} 
where the $\Gamma^j_{ki}=\Gamma^j_{ik}$ are called Christoffel symbols and related with $g_{ij}$ from the relations
\begin{equation}
\partial_l g_{ik}=\sum^{N}_{n=1}g_{nl}\Gamma^{n}_{ki}+\sum^{N}_{n=1}g_{kn}\Gamma^{n}_{il}  
\end{equation}
and
\begin{equation}
\Gamma^{i}_{kl}=\frac{1}{2}\sum^{N}_{n=1}(g_{in})^t(\partial_k g_{nl}+\partial_lg_{kn}-\partial_ng_{kl}).
\end{equation}
\\
\textbf{Note.}\\
The Beltrami operator of the first kind for one function is by notation
\begin{equation}
\Delta_1\Phi:=\frac{1}{2}\Delta_1(\Phi,\Phi)=\sum^{N}_{i,j=1}(g_{ij})^t\frac{\partial\Phi}{\partial u_i}\frac{\partial \Phi}{\partial u_j}=\sum^{N}_{i,j=1}(g_{ij})^t\partial_i \Phi \partial_j\Phi. 
\end{equation}
Hence in $S$ the Schrodinger's equation with potential $V=V(x_1,x_2,\ldots,x_N,t)$, (relaxed from the physical symbols which we may take them equal to 1, without loss of the generality of the problem) read as 
\begin{equation}
i\frac{\partial\Psi(x_1,x_2,\ldots,x_N,t)}{\partial t}=-\Delta_{2,x}\Psi(x_1,x_2,\ldots,x_N,t)+V\Psi(x_1,x_2,\ldots,x_N,t).
\end{equation}
The $x$ index in the $\Delta$ derivative in (11) means that only the $x$'s are differentiated and the $t$'s is parameter (the Beltrami derivative is for the space $S$).\\
\\
\textbf{Definition 3.}\\
The space $S$ in which all the $S_k(x_1,x_2,\ldots,x_N)$, $k=1,2,\ldots,N$ are homogeneous of degree $\mu$, we call it $S^{N}_{\mu}$ ($\mu-$Homogeneous Space).\\
\\
\textbf{Definition 4.}\\
We call $\textbf{F}^{N}_{\mu}$, the space of all real smooth functions $(f,g)$ such that  
\begin{equation}
\lim_{r\rightarrow \{0,+\infty\}}r^{(N-1)\mu}\left(g(r)f'(r)-f(r)g'(r)\right)=0
\end{equation}
and
$$
(\mu-1)(N-1)\int^{\infty}_{0}r^{(N-1)\mu-1}\left(F'(r)G_1(r)-F_1(r)G_1'(r)\right)dr=0,\eqno{(12.1)}
$$
where $N$ is the dimension of $S$.\\
\\
\textbf{Note.}\\ 
Observe that in $F^{N}_{1}$ space we have only condition (12), since then (12.1) is always true.\\
\\
\textbf{Definition 5.}\\
We will call the function $A=A(x^a)=A(x_1,x_2,\dots,x_N)$ harmonic if 
\begin{equation}
\Delta_2A(x^a)=0.
\end{equation}
\\

We note here that with $x^a$ we mean $x^a=\{x^a\}=\{x_1,x_2,\ldots,x_N\}$, for the $N$ first values. If $A$ has more than $N$ cordinates, say $x=\{x_1,x_2,\ldots,x_N,x_{N+1},\ldots,x_{M}\}$, then we will write $x=\{x^a,x_{N+1},\ldots,x_{M+1}\}$. Hence the symbols $a$, $x^{a}$ are protected.

\section{Main tasks of the article}   

We will solve the Schroedinger's equation (11) for the potentials 
\begin{equation}
V(x^a,t)=\frac{\sum^{\infty}_{\lambda=1}c^{*}_{\lambda}\lambda e^{-it\lambda}A_{\lambda}(x^a)}{\sum^{\infty}_{\lambda=1}c^{*}_{\lambda} e^{-it\lambda}A_{\lambda}(x^a)+\sum^{\infty}_{\lambda=1}c_{\lambda}\phi_{\nu,\lambda}(\Phi_0)e^{-it\lambda}}, 
\end{equation}
where $\phi_{\nu,\lambda}(x)$ are specific (null) functions and $A_{\lambda}(x^a)=A_{\lambda}(x_1,x_2,\ldots,x_N)$ are arbitrary harmonic function of the space $S$. Also we have set      
\begin{equation}
\Phi_0=\Phi_0(x_1,x_2,\ldots,x_N)=\sqrt{\sum^{N}_{k=1}S_{k}(x_1,x_2,\ldots,x_N)^2} . 
\end{equation}
The function $\Phi_0$ is what we call radial distance and its use is very important, since in many cases reduces considerably the complexity of the problem. Also in an Eucledian space, $\Phi_0$ have the very well known distance meaning used in central potentials.\\    
We also show that, in every space $S$ exists an infinite class of harmonic functions $A(x^a)$, which will construct  them (however we don't find all of them). In Eucledian spaces such functions have been found in full generality (see Note 3 in page 17 below), but in the cases of arbitrary spaces are hard to find.\\
Also in spaces $S^{N}_{\mu}$ over $\textbf{F}^{N}_{\mu}$, the potentials $V(x^a,t)$ are not only exactly solvable but produce together with the Beltrami derivative $\Delta_2$ hamiltonians: 
$$
H(.)=-\Delta_{2,x}(.)+V(x^a,t)(.)\textrm{, }
$$
which are self-adjoint i.e. Hermitian. The concept of self-adjointnes of the hamiltonian $H$ is fundamental because gives a physical meaning to the quantum system. Without this one can not proceed further.\\
We give also several applications and solutions of certain equations in the most general Riemann spaces $S$. Note also the very interesting appearance of Bessel functions of the first and second kind $J$ and $Y$, which are related with the $\Delta_2$ Beltrami operators. Thsese functions ($J$, $Y$), are generated in the middle of nowhere, when we consider the  eigenvectors and eigenvalues of the Beltrami oparators in curved spaces. Examples of solutions of equation (11) in spaces $S^{N}_{\mu}$ over $\textbf{F}^{N}_{\mu}$, with the Hamiltonian self adjoint and radial potentials of the form 
$$
V=V(x^a,t)=\left(-\frac{C}{x}\frac{\partial }{\partial x}\log\left(\sum^{\infty}_{\lambda=1}c_{\lambda}e^{-it\lambda}\phi_{\nu,\lambda}(x)\right)\right)_{x=\Phi_0(x^a)}=
$$
\begin{equation}
=\left(-\frac{C}{x}\frac{\partial }{\partial x}\log\left(g(x,t)\right)\right)_{x=\Phi_0(x^a)},
\end{equation}
where $\phi_{\nu,\lambda}(x)=J_{\nu}\left(\sqrt{\lambda}x\right)/x^{\nu}$, $C-$constant and $\nu$ real are given.\\We also show that for the PDE
\begin{equation}
i\partial_tg(x^a,t)=-\Delta_2g(x^a,t),
\end{equation} 
a class of solutions is
\begin{equation}
g(x^a,t)=\sum^{\infty}_{\lambda=1}c_{\lambda}e^{-it\lambda}\phi_{\nu,\lambda}\left(\Phi_0(x^a)\right), 
\end{equation}
where
\begin{equation}
\phi_{\nu,\lambda}(x)=x^{-\nu}\left(c_1J_{\nu}(\sqrt{\lambda}x)+c_2Y_{\nu}(\sqrt{\lambda}x)\right).
\end{equation}
It is worth to mention that in Eucledian spaces $\bf R\rm$, $\bf R^2\rm$ and $\bf R^3\rm$ relation (16) takes the form:
\begin{equation}
V=V(x^a,t)=\left(\frac{C}{x}\frac{\partial}{\partial x}\log\left(\sum^{\infty}_{\lambda=1}c_{\lambda}e^{-it\lambda}\phi_{\nu,\lambda}(x)\right)\right)_{x=\Phi_{1,2,3}},
\end{equation}
with
$$
\Phi_1=\Phi_0=x_1\textrm{ , } \Phi_2=\Phi_0=\sqrt{x_1^2+x_2^2}\textrm{ , }\Phi_3=\Phi_0=\sqrt{x_1^2+x_2^2+x_3^2}
$$
and $\nu=-1/2,0,1/2$ respectively. In the case of $S=\textbf{R}$, we have $\Phi_0=x_1=x$ and hence we solve
\begin{equation}
i\partial_t\Psi(x,t)=-\partial^2_x\Psi(x,t)+V(x,t)\Psi(x,t),
\end{equation}
for the potential (20) in full generality. This equation admits exact solution
\begin{equation}
\Psi(x,t)=\sum^{\infty}_{p=1}\left(\frac{2\sqrt{x}}{J_{1/2}(\lambda_{p})^2}\int^{1}_{0}f(y)y^{1/2}J_{-1/2}\left(\lambda_{p}y\right)dy\right)e^{-i\lambda_{p}^2t}J_{-1/2}\left(\lambda_{p}x\right),
\end{equation}
where the numbers $\lambda_p$ are the zeros of $J_{-1/2}(x)=0$ in $x>0$ and  
\begin{equation}
V(x,t)=\frac{C}{x}\partial_x\left(\log(g(x,t))\right)\textrm{, }g(x,t)=\sum^{\infty}_{p=1}c_{p}e^{-it\lambda_p^2}\phi_{-1/2,\lambda_p^2}(x),
\end{equation}
with $u(1,t)=0$, $t>0$ and $u(x,0)=f(x)$. Note that when $c_2=0$ and $r_{\nu,p}$ are the solutions of $J_{\nu}(x)=0$, then $x^{\nu+1/2}\phi_{\nu,r_{\nu,p}^2}(x)$ is base of $L[0,1]$.

\section{Properties of Beltrami differential oparators and Schrodinger's equation}

\textbf{Proposition 1.}
\begin{equation}
\Delta_1(\Phi\Psi)=\Phi^2(\Delta_1\Psi)+(\Delta_1\Phi)\Psi^2+\Phi\Psi\Delta_1(\Phi,\Psi).
\end{equation}
\\
\textbf{Proof.}\\
The result follows from the differentiation of the product of two functions and the fact that $g_{ij}$ is symmetric:
$$
\Delta_1(\Phi\Psi)=\sum^{N}_{i,j=1}(g_{ij})^t\partial_i(\Phi\Psi)\partial_j(\Phi\Psi)=
$$
$$\sum^{N}_{i,j=1} (g_{ij})^t[(\Psi(\partial_i\Phi)+\Phi(\partial_i\Psi))((\partial_j\Phi)\Psi+\Phi(\partial_j\Psi))]=
$$
$$
=\sum^{N}_{i,j=1}(g_{ij})^t[\Psi^2(\partial_i\Phi)(\partial_j\Phi)+\Phi^2(\partial_i\Phi)(\partial_j\Psi)+\Phi\Psi(\partial_i\Phi)(\partial_j\Psi)+\Phi\Psi(\partial_j\Phi)(\partial_i\Psi)]=
$$
$$
=\Phi^2(\Delta_1\Psi)+\Psi^2(\Delta_1\Phi)+\Phi \Psi \Delta_1(\Phi,\Psi).
$$ 
\\  
\textbf{Proposition 2.}
\begin{equation}
\Delta_1(\Phi \Psi,Z)=\Phi\Delta_1(\Psi,Z)+\Psi\Delta_1(\Phi,Z).
\end{equation}
\\
\textbf{Proof.}
$$
\Delta_1(\Phi \Psi,Z)=2\sum^{N}_{i,j=1}(g_{ij})^t\partial_i(\Phi\Psi)\partial_j Z=2\sum^{N}_{i,j=1}(g_{ij})^t\left(\Psi\partial_i\Phi+\Phi\partial_i\Psi\right)\partial_jZ=
$$
$$
=2\Phi \sum^{N}_{i,j=1}(g_{ij})^t(\partial_i\Psi)(\partial_jZ)+2\Psi \sum^{N}_{i,j=1}(g_{ij})^t(\partial_i\Phi)(\partial_jZ)
$$
and the result follows from the definition of $\Delta_1(.,.)$.\\
\\

One can observe from Proposition 2 that Beltrami differential operator for products, obey the same rule as the classical differential operator $\frac{d}{dx}$ of functions of one variable. For example set $\Phi=\Psi$ in (25), we get:
\begin{equation}
\Delta_1(\Phi^2,Z)=2\Phi\Delta_1(\Phi,Z) .
\end{equation} 
From Propositions 1 and 2 we get by induction.\\
\\
\textbf{Proposition 3.}\\
If $n=1,2,3,\ldots$, then
\begin{equation}
\Delta_1(\Phi^n,\Psi)=n\Phi^{n-1}\Delta_1(\Phi,\Psi).
\end{equation} 
\\

The semi-linear property which is easy someone to see is
\begin{equation}
\Delta_1(\Phi+\Psi,Z)=\Delta_1(\Phi,Z)+\Delta_1(\Psi,Z).
\end{equation}
Also if $f$ is a function such that 
\begin{equation}
f(z)=\sum^{\infty}_{n=1}c_nz^n,
\end{equation}
then\\
\\
\textbf{Theorem 1.}
\begin{equation}
\Delta_1(f(\Phi),Z)=f'(\Phi)\Delta_1(\Phi,Z).
\end{equation}
\\
\textbf{Proof.}\\
It follows from Proposition 3 and the semi-linear property.\\
\\

Theorem 1 will help us to evaluate the second Beltrami operator $\Delta_2(f(\Phi))$ of the one variable function $f(x)$ on a general scalar $\Phi$.\\
\\
\textbf{Proposition 4.}
\begin{equation}
\Delta_2(\Phi\Psi)=\Psi\Delta_2(\Phi)+\Phi\Delta_2(\Psi)+\Delta_1(\Phi,\Psi).
\end{equation}
\\
\textbf{Proof.}\\
From the relations
$$\frac{\partial^2(\Phi\Psi)}{\partial x\partial y}=\frac{\partial^2\Phi}{\partial x\partial y}+\frac{\partial^2\Psi}{\partial x \partial y}+\frac{\partial\Phi}{\partial x}\frac{\partial \Psi}{\partial y}+\frac{\partial \Phi}{\partial y}\frac{\partial \Psi}{\partial x},$$   
$$\frac{\partial(\Phi\Psi)}{\partial x}=\frac{\partial \Phi}{\partial x}\Psi+\frac{\partial \Psi}{\partial x}\Phi$$
and the  definitions of 1 and 2-Beltrami derivatives, the proof easily follows.\\
\\
\textbf{Proposition 5.}\\
For $n=2,3,...$
\begin{equation}
\Delta_2(\Phi^n)=n\Phi^{n-1}\Delta_2(\Phi)+n(n-1)\Phi^{n-2}\Delta_1(\Phi).
\end{equation}
\\
\textbf{Proof.}\\
Set $\Phi=\Psi$ in (31) then
\begin{equation}
\Delta_2(\Phi^2)=2\Phi\Delta_2(\Phi)+\Delta_1(\Phi,\Phi).
\end{equation} 
Also
$$\Delta_2(\Phi^3)=\Delta_2(\Phi\Phi^2)=\Phi^2\Delta_2(\Phi)+\Phi\Delta_2(\Phi^2)+\Delta_1(\Phi^2,\Phi).$$
But from Theorem 1. and (19) we have 
$$\Delta_2(\Phi^3)=\Phi^2\Delta_2(\Phi)+\Phi(2\Phi\Delta_2(\Phi)+\Delta_1(\Phi,\Phi))+2\Phi\Delta_1(\Phi,\Phi)=
$$
$$=3\Phi^2\Delta_2(\Phi)+3\Phi\Delta_1(\Phi,\Phi).$$
The result as someone can see follows easy from the above propositions and theorems by induction.\\
\\
From the linearity of the 2-Beltrami operator and (32) we get\\
\\
\textbf{Theorem 2.}\\
If $f$ is a single variable function and analytic around 0  we have
\begin{equation}
\Delta_2(f(\Phi))=f'(\Phi)\Delta_2(\Phi)+f''(\Phi)\Delta_1(\Phi).
\end{equation}
\\  
\textbf{Notes.}\\
1) From Theorem 2 we get that if for some $\Phi$ in some space holds 
\begin{equation}
\Delta_1(\Phi)=\Delta_2(\Phi)=0 ,
\end{equation}
then for every single variable function $f$ analytic in a open set, containing the origin, we have
\begin{equation} 
\Delta_2(f(\Phi))=0.
\end{equation}
2) Set 
$$\left|u\right|:=\sqrt{\sum^{N}_{k=1}u^2_k}$$
and
$$\left\|u\right\|^2:=\sum^{N}_{i,j=1}(g_{ij})^tu_iu_j,$$
then
\begin{equation}
\Delta_1(\left|u\right|)=\frac{\left\|u\right\|^2}{\left|u\right|^2}.
\end{equation}
Because
$$
\Delta_1(\left|u\right|)=\sum^{N}_{i,j=1}(g_{ij})^t\frac{\partial \left|u\right|}{\partial u_i}\frac{\partial \left|u\right|}{\partial u_j}=\sum^{N}_{i,j=1}(g_{ij})^t\frac{\partial \sqrt{\sum^{N}_{k=1}u^2_k}}{\partial u_i}\frac{\partial\sqrt{\sum^{N}_{k=1}u^2_k}}{\partial u_j}=
$$
$$
=\sum^{N}_{i,j=1}(g_{ij})^t2u_i 2 u_j\frac{1}{4\left|u\right|^2}=\frac{\left\|u\right\|^2}{\left|u\right|^2}.$$
Hence one can also see that
\begin{equation}
\Delta_1[f(\left|u\right|),g\left(\left|u\right|\right)]=2f'(\left|u\right|)g'\left(\left|u\right|\right)\frac{\left\|u\right\|^2}{\left|u\right|^2}.
\end{equation}
\begin{equation}
\Delta_2(f(\left|u\right|))=f'(\left|u\right|)\Delta_2(\left|u\right|)+f''(\left|u\right|)\frac{\left\|u\right\|^2}{\left|u\right|^2}.
\end{equation}
3) Theorems 1,2 can be used for calculation of the Beltrami derivatives. One can easily see that also hold the following relations
$$
\Delta_1(u^{\lambda})=(g_{\lambda\lambda})^t.
$$
$$
\Delta_2\left(u_\lambda\right)=-\sum^{N}_{i,j=1}(g_{ij})^t\Gamma^{\lambda}_{ij}=T_{\lambda}.
$$
If $M\leq N$, then for all smooth functions $f$ and $\Phi_k$ we have 
$$
\Delta_1\left(\sum^{M}_{k=1}f(\Phi_k)\right)=
$$
$$
=\sum^{M}_{k=1}\left(f'(\Phi_k)\right)^2\Delta_1(\Phi_k)+2\sum^{M}_{(k<\mu),k,\mu=1}f'(\Phi_k)f'(\Phi_{\mu})\Delta_1(\Phi_k,\Phi_{\mu}).\eqno{:(a)}
$$
Hence for example if we have to evaluate 'say' $\Delta_2(\log(u_2+u_3)+1/u_1)$ in a arbitrary 3-dimensional space, then
$$\Delta_2(\log(u_2+u_3)+u_1^{-1})=$$
$$=\frac{1}{u_2+u_3}\Delta_2(u_2+u_3)+\frac{-1}{(u_2+u_3)^2}\Delta_1(u_2+u_3)+\frac{-1}{u_1^2}\Delta_2(u_1)+\frac{2}{u_1^3}\Delta_1(u_1)=$$
$$
=\frac{T_2+T_3}{u_2+u_3}-\frac{\Delta_1(u_2)+\Delta_1(u_3)+2\Delta_1(u_2,u_3)}{(u_2+u_3)^2}-\frac{T_1}{u^2_1}+\frac{2(g_{11})^t}{u^3_1}=
$$
$$
=\frac{T_2+T_3}{u_2+u_3}-\frac{(g_{22})^t+(g_{33})^t+2(g_{23})^t}{(u_2+u_3)^2}-\frac{T_1}{u^2_1}+\frac{2(g_{11})^t}{u^3_1}.
$$
\\

Now we considering the following parametrization 
$$
A=\left\{x_l : x_l=x_l(p) \textrm{, and } \Phi(x_1(p),x_2(p),\ldots,x_N(p))=p, p\in\bf R\rm \right\}
$$ 
and $f$ such that $\Delta_2(f(\Phi))_A=0$, from Theorem 2 we have  
$$
\Delta_2(f(\Phi))_A=f'(\phi)(\Delta_2(\Phi))_A+f''(\phi)(\Delta_1(\Phi))_A=0,
$$   
which is an ordinary differential equation in a single variable (here the variable is $\phi$), which the solution is 
\begin{equation}
f(\phi)=\int\exp\left[-\int\frac{\Delta_2(\Phi)_A}{\Delta_1(\Phi)_A}d\phi\right]d\phi. 
\end{equation}
Hence we have the next\\
\\
\textbf{Theorem 4.}\\
Let $\Phi=\Phi(x_1,x_2,\ldots,x_N)$ be a function of a certain differentiable class. If the parametrization $A$ is such that 
$$\Phi(\left\{x_l(p)\right\})=p.$$ 
Then        
\begin{equation}
\Delta_2\left(\int^{\Phi(x_1,x_2,\ldots,x_N)}\exp\left[-\int\frac{\Delta_2(\Phi)_A}{\Delta_1(\Phi)_A}d\phi\right]d\phi\right)_A=0 .
\end{equation}
\\
\textbf{Examples.}\\
Consider the 4-dimensional space
$$
\overline{S}(u_1,u_2,u_3,u_4)=\{u_1,u_1+u_2,u_1u_3,u_4\}.
$$
Then
$$
\overline{e}_i=\frac{\partial \overline{S}}{\partial u_i}
$$
and 
$$ 
g_{ij}=\left\langle \overline{e}_i,\overline{e}_j\right\rangle_{reg}. 
$$
i) Set $$\Phi(u_1,u_2,u_3,u_4)=\sqrt{u_1^2+u_2^2+u_3^2+u_4^2}.$$ 
Then one parametrization is $u_2=u_3=u_4=p$ and 
$$
u_1=h(p)=\sqrt{-2}p . 
$$
The function $f$ is 
$$
f(\phi)=\int\exp\left[-\int\frac{\Delta_2(\Phi)_A}{\Delta_1(\Phi)_A}d\phi\right]d\phi=
$$ 
$$
=\int\exp\left[\frac{3\sqrt{2} -4i}{12\sqrt{2}-6i}\left(2i\arctan\left(\frac{6\sqrt{2} \phi^2}{1+3\phi^2}\right)+\log\left(1+6\phi^2+81 \phi^4\right)\right)\right]d\phi 
$$ 
and 
$$
A_1=\left\{\sqrt{-2}p,p,p,p\right\}, p\in\bf R\rm .
$$ 
If one set the above values in the second order Beltrami derivative then 
$$
\Delta_2(f(\Phi))_{A_1}=0 . 
$$
ii) Set
 $$
 \Phi(u_1,u_2,u_3,u_4)=\frac{u_1^2}{\sqrt{u_2+u_3+u_4}}.
 $$ 
Then one parametrization is $u_2=u_3=u_4=p$, $u_1=\sqrt{3}p$. 
The function $f$ is 
$$
f(\phi)=-\frac{{}_2F_1\left[-\frac{1}{2},-\frac{18}{29+7 \sqrt{3}}-\frac{3 \sqrt{3}}{29+7 \sqrt{3}};\frac{1}{2};-\left(58+14 \sqrt{3}\right) \phi^2\right]}{\phi} 
$$ 
and 
$$
A_2=\left\{\sqrt{3}p,p,p,p\right\}, p\in\bf R\rm . 
$$
Then indeed we get
$$
\Delta_2(f(\Phi))_{A_2}=0 . 
$$
\\
\textbf{Theorem 5.}\\
Let $P$ be any point of a metric space $S$. Let also that $S$ is described by the vector 
$$
\overline{OP}=\overline{S}=\left\{S_1(x_1,x_2,\ldots,x_N),S_2(x_1,x_2,\ldots,x_N),\ldots,S_N(x_1,x_2,\ldots,x_N)\right\}.\eqno{(41.1)}
$$ 
Then it holds
\begin{equation}
\Delta_2f\left(\Phi_0(x_1,x_2,\ldots,x_N)\right)=\left[x^{-{(N-1)}}\frac{d}{dx}\left(x^{N-1}f'(x)\right)\right]_{x=\Phi_0(x_1,x_2,\ldots,x_N)}
\end{equation}
and
\begin{equation}
\Delta_1(f(\Phi_0))=f'(\Phi_0)^2 , 
\end{equation}
where
\begin{equation}
\Phi_0(x_1,x_2,\ldots,x_N)=\sqrt{\sum^{N}_{k=1}S_k(x_1,x_2,\ldots,x_N)^2} .
\end{equation}
\\

We give the idea of how we arrived to this result. The calculations are not proper but discribe the idea:\\
From (34) it holds   
$$
\frac{\Delta_2(f(\Phi))}{\Delta_1(\Phi)}=f'(\Phi)\frac{\Delta_2(\Phi)}{\Delta_1(\Phi)}+f''(\Phi).
$$ 
Or if we use the parametrization of Theorem 4
$$
e^{\int{\frac{\Delta_2(\Phi)}{\Delta_1(\Phi)}d\Phi}}\frac{\Delta_2(f(\Phi))}{\Delta_1(\Phi)}=\frac{d}{d\Phi}\left(e^{\int\frac{\Delta_2(\Phi)}{\Delta_1(\Phi)}d\Phi}f'(\Phi)\right)
$$
Or
\begin{equation}
\int e^{-\int\frac{\Delta_2(\Phi)}{\Delta_1(\Phi)}d\Phi}\int e^{\int{\frac{\Delta_2(\Phi)}{\Delta_1(\Phi)}}d\Phi}\frac{\Delta_2(f(\Phi))}{\Delta_1(\Phi)}d\Phi=f(\Phi).
\end{equation}
Now, if $\Phi$ is that of (44), then  $\Delta_2(f(\Phi))$, $\Delta_1(\Phi)$, $\Delta_2(\Phi)$ are functions of $\Phi$ and we have  
\begin{equation}
\Delta_1(\Phi)=1
\end{equation}
and
\begin{equation}
\Delta_2(\Phi)=\frac{N-1}{\sqrt{\sum^{N}_{k=1}S_k(x_1,x_2,\ldots,x_N)^2}}=\frac{N-1}{\Phi} .
\end{equation}
From (44),(45),(46),(47) and setting 
$$
H(x)=\frac{\Delta_2(f(\Phi))}{\Delta_1(f(\Phi))}, 
$$ 
we arrive to 
$$
\int x^{-(N-1)}\left(\int x^{(N-1)}H(x)dx\right)dx=f(x).
$$
Hence solving with respect to $H(x)$ this last equation, we find the value of $\Delta_2f(\Phi)$, where  
$$
\Phi=\Phi_0=\sqrt{\sum^{N}_{n=1}S_{n}(x_1,x_2,\ldots,x_N)^2}.
$$
\\
\textbf{Theorem 6.}\\
Consider the $N-$dimensional metric space $S$ (as in Theorem 5) and $\nu=(N-2)/2$. Then the PDE
\begin{equation}
\partial_t U(x^a,t)=\Delta_{2,x}U(x^a,t),
\end{equation}
admits solution
$$
U(x_1,x_2,\ldots,x_N,t)=U(x^a,t)
=\frac{C_1}{\Phi_0(x^a)^{\nu}}\sum^{\infty}_{\lambda=1}c_{\lambda}e^{-\lambda t}J_{\nu}\left(\sqrt{\lambda}\Phi_0(x^a)\right)+
$$
\begin{equation}
+\frac{C_2}{\Phi_0(x^a)^{\nu}}\sum^{\infty}_{\lambda=1}c_{\lambda}e^{-\lambda t}Y_{\nu}\left(\sqrt{\lambda}\Phi_0(x^a)\right) , 
\end{equation}
where $J_{\nu}$ and $Y_{\nu}$ are the usual Bessel functions of the first and second kind of order $\nu$ and 
$$
\Phi_0(x^a)=\sqrt{\sum^{N}_{k=1}S_k(x^a)^2} . 
$$
\\
\textbf{Proof.}\\
Set $X_{\lambda}(x^a)=X_{\lambda}(\Phi_0)$ and observe from (42) that the DE
$$
\Delta_2X_{\lambda}(x^a)=-\lambda X_{\lambda}(x^a)
$$
have solution
$$
X_{\lambda}(x^a)=y_{\lambda}(\Phi_0),
$$ 
where
$$
y_{\lambda}(x)=C_1x^{-\nu}J_{\nu}(\sqrt{\lambda}x)+C_2x^{-\nu}Y_{\nu}(\sqrt{\lambda}x). 
$$
This lead us to the desired result.\\
\\
\textbf{Notes.}\\ 
1) The main idea of Theorem 6 remains the same if we take instead of $\lambda$ an arbitrary sequence $\lambda_p$. Then the summation will be with respect to $p$.\\ 
2) The function
$$
U_1(x_1,x_2,\ldots,x_N,t)=U(x^a,t)
=\frac{C_1}{\Phi_0(x^a)^{\nu}}\sum^{\infty}_{\lambda=1}c_{\lambda}e^{-i\lambda t}J_{\nu}\left(\sqrt{\lambda}\Phi_0(x^a)\right)+
$$
\begin{equation}
+\frac{C_2}{\Phi_0(x^a)^{\nu}}\sum^{\infty}_{\lambda=1}c_{\lambda}e^{-i\lambda t}Y_{\nu}\left(\sqrt{\lambda}\Phi_0(x^a)\right) , 
\end{equation}
satisfies the equation
\begin{equation}
i\partial_tU_1(x^a,t)=-\Delta_{2,x}U_1(x^a,t).
\end{equation}
3) For any function $f(x)$ of one variable and $\Phi_0=\sqrt{\sum^{N}_{k=1}S_k(x^a)^2}$, holds
\begin{equation}
\Delta_{2,x}f\left(\Phi_0\right)=f''\left(\Phi_0\right)+\frac{N-1}{\Phi_0}f'\left(\Phi_0\right).
\end{equation}
\\    
\textbf{Theorem 7.}\\
Consider the $N-$dimensional metric space $S$ (as in Theorem 5). The PDE
\begin{equation}
\Delta_2U(x_1,x_2,\ldots,x_N)=-\lambda U(x_1,x_2,\ldots,x_N)\textrm{, }\lambda>0,
\end{equation}
admits solution
$$
U(x_1,x_2,\ldots,x_N)
=\frac{C_1}{\Phi_0(x_1,x_2,\ldots,x_N)^{N/2-1}}J_{N/2-1}\left(\sqrt{\lambda}\Phi_0(x_1,x_2,\ldots,x_N)\right)+
$$
\begin{equation}
+\frac{C_2}{\Phi_0(x_1,x_2,\ldots,x_N)^{N/2-1}}Y_{N/2-1}\left(\sqrt{\lambda}\Phi_0(x_1,x_2,\ldots,x_N)\right) . 
\end{equation} 
\\
\textbf{Proof.}\\
Easy.\\
\\
\textbf{Theorem 8.}\\
In the space $S$, we set the Beltrami-D'Alembert wave operator to be 
\begin{equation}
\Pi_2\equiv\Delta_{2}-\frac{1}{c^2}\frac{\partial^2}{\partial t^2},
\end{equation}
where the $\Delta_2$ is defined in the space $S$ as in (7). Then the equation 
$$
\Pi_2 U(x^a,t)=0 , 
$$
where $x^a=\left\{x_1,x_2,\ldots,x_N\right\}$
admits solution
\begin{equation}
U(x^a,t)=\sum^{\infty}_{\lambda=1}B_{\lambda}e^{itc\sqrt{\lambda}}U_{\lambda}(x^a) . 
\end{equation} 
The function $U_{\lambda}(x^a)=U(x^a)$, $U(x^a)$ is that of (54).\\
\\
\textbf{Proof.}\\
Easy\\
\\
\textbf{Theorem 9.}\\
Let $S^{+}$ be the space produced as
$$
\overline{S}^{+}(x^a,x_{N+1})=
$$
$$
=\left\{S_1(x^a,x_{N+1}),S_2(x^a,x_{N+1}),\ldots,S_N(x^a,x_{N+1}),S_{N+1}(x^a,x_{N+1})\right\} . 
$$
Then if
$$
U_{\lambda}(x_1,x_2,\ldots,x_N,x_{N+1})=
$$
$$
=U_{\lambda}(x^a,x_{N+1})
=\frac{C_1}{\Phi_0(x^a,x_{N+1})^{N/2-1}}J_{N/2-1}\left(\sqrt{\lambda}\Phi_0(x^a,x_{N+1})\right)+
$$
$$
+\frac{C_2}{\Phi_0(x^a,x_{N+1})^{N/2-1}}Y_{N/2-1}\left(\sqrt{\lambda}\Phi_0(x^a,x_{N+1})\right),  
$$
is that of relation (54) (with $N+1$ coordinates) and
$$
\Phi_0(x^a,x_{N+1})=\sqrt{\sum^{N}_{k=1}S^{+}_k(x^a,x_{N+1})^2},
$$
then the function
$$
U^{+}=U^{+}(x_1,x_2,\ldots,x_N,x_{N+1})=
$$
\begin{equation}
=U^{+}(x^a,x_{N+1})=\sum^{\infty}_{\lambda=1}c_{\lambda}e^{\sqrt{\lambda}S_{N+1}(x^a,x_{N+1})}U_{\lambda}(x^a,x_{N+1}),
\end{equation}
have 2-Beltrami derivative 
\begin{equation}
\Delta_2\left(U^{+}(x_1,x_2,\ldots,x_N,x_{N+1})\right)=0,
\end{equation}
where $\nu=(N+1-3)/2$. Here for no confusion the second Beltrami derivative reffers to all coordinates $\{x^a,x_{N+1}\}=\{x_1,x_2,\ldots,x_N,x_{N+1}\}$.\\
\\
\textbf{Proof.}\\
Let $x^a$, $x_{N+1}$ as defined in the state of the theorem. We will use the following acceptable change of variables:  
$$
x_j=f_j\left((x^a)',x'_{N+1}\right) , j=1,2,\ldots,N+1 , 
$$
such that:  
$$
S_k(x^a,x_{N+1})=S_k\left((x^a)',0\right) , \textrm{ for } 1\leq k\leq N \textrm{ and } 1\leq a\leq N \eqno{: (\nu_1)} 
$$
and
$$
S_{N+1}(x^a,x_{N+1})=icx'_{N+1}. \eqno{:(\nu_2)}
$$ 
Also let  $\Phi_0=\sqrt{\sum^{N}_{k=1}S_k\left(x^a\right)^2}$ considers the values of $S_k$ for $k=1,\ldots,N$ only. Then it holds  
$$
\Pi_2\left(\sum^{\infty}_{\lambda=1}c_{\lambda}e^{ic\sqrt{\lambda}x'_{N+1}}U_{\lambda}(x'_1,x'_2,\ldots,x'_N)\right)=0 
$$ 
and also
$$
\Delta_2U^{+}(x^a,x_{N+1})
=\Delta_2\left(\sum^{\infty}_{\lambda=1}c_{\lambda}e^{\sqrt{\lambda} S_{N+1}(x^a,x_{N+1})}U_{\lambda}(x_1,x_2,\ldots,x_N,x_{N+1})\right) .
$$
But from the change of variables $(\nu_1)$ and $(\nu_2)$ 
we have
$$
\Delta_2U^{+}(x^a,x_{N+1})=\Delta_2\left(\sum^{\infty}_{\lambda=1}c_{\lambda}e^{i\sqrt{\lambda} c x'_{N+1}}U_{\lambda}(x'_1,x'_2,\ldots,x'_N)\right)=
$$
$$
=\Pi_2\left(\sum^{\infty}_{\lambda=1}c_{\lambda}e^{ic\sqrt{\lambda}x'_{N+1}}U_{\lambda}(x'_1,x'_2,\ldots,x'_N)\right)=0 , 
$$
according to Theorem 8. Hence we complete the proof.\\
\\
\textbf{Note.}\\ The above theorem show us that in every metric space are attached the harmonic functions (57), which by Theorems 6,7 are related with the Bessel functions $J_{\nu}(x)$ and $Y_{\nu}(x)$.\\
\\ 
\textbf{Theorem 10.}\\
Let $S^{++}$ be a metric space with 
$$
\overline{S}^{++}(x_1,x_2,\ldots,x_N,x_{N+1},x_{N+2})=\overline{S}^{++}(x^a,x_{N+1},x_{N+2})=
$$
$$
=\left\{S_1(x^a),S_{2}(x^a),\ldots,S_{N}(x^a),x_{N+1},x_{N+2}\right\}
$$
and 
$$
\Phi_0=\Phi_0(x^a)=\sqrt{\sum^{N}_{k=1}S^2_k(x^a)}.
$$ 
Then the equation
$$
\Delta_2U^{++}-N\sum^{N+2}_{k,m=N+1}\epsilon_{k,m}\frac{\partial^2 U^{++}}{\partial x_{k}\partial x_m}=0,
$$
where $\epsilon_{N+1,N+1}=\epsilon_{N+2,N+2}=1$ and $\epsilon_{N+1,N+2}=\epsilon_{N+2,N+1}=-1/2$, admits solution 
$$
U^{++}=\sum^{\infty}_{k=1}A_{k}e^{i x_{N+1}\lambda_k}e^{i x_{N+2}\lambda_k}\frac{J_{N/2-1}\left(\Phi_0\lambda_k\right)}{\Phi^{N/2-1}_0}
$$
and $\Delta_2$ is refering to all coordinates $\{x_1,x_2,\ldots,x_N,x_{N+1},x_{N+2}\}$.\\ 
\\
\textbf{Theorem 11.}\\
Let $S$ be the $3-$dimensional flatten metric space. Set $x^a=\left\{x_1,x_2,x_3\right\}$ and let $A_{\lambda}=A_{\lambda}(x^a)$ such that 
\begin{equation}
\Delta_2\left(A_{\lambda}(x^a)\right)=0. 
\end{equation}
Moreover let 
$$
\phi_{\lambda}(x)=\frac{\sqrt{2}}{\sqrt{\pi}\lambda^{1/4}x}\left(C_1\sin(\sqrt{\lambda}x)+C_2\cos(\sqrt{\lambda} x)\right)
$$
and 
$\phi_{\lambda}=\phi_{\lambda}(\Phi_0)$, with  $\Phi_0=\sqrt{x_1^2+x_2^2+x_3^2}$.\\ 
If
\begin{equation}
V(x^a,t)=\frac{\sum^{\infty}_{\lambda=1}c^{*}_{\lambda}\lambda e^{-it\lambda}A_{\lambda}(x^a)}{\sum^{\infty}_{\lambda=1}c^{*}_{\lambda}e^{-it\lambda}A_{\lambda}(x^a)+\sum^{\infty}_{\lambda=1}c_{\lambda}\phi_{\lambda}(\Phi_0)e^{-it\lambda}},  
\end{equation} 
then a solution of the Schrodinger's equation
\begin{equation}
i\partial_t\Psi(x^a,t)=-\Delta_{2,x}\Psi(x^a,t)+V(x^a,t)\Psi(x^a,t),
\end{equation}
is
\begin{equation}
\Psi(x^a,t)=\sum^{\infty}_{\lambda=1}c_{\lambda}\phi_{\lambda}(\Phi_0)e^{-it\lambda}+\sum^{\infty}_{\lambda=1}c^{*}_{\lambda}e^{-it\lambda}A_{\lambda}(x^a).
\end{equation}
\\  
\textbf{Proof.}\\
Let $\Psi=\Psi(x^a,t)=U(x^a,t)+G(x^a)$, where $U(x^a,t)=\sum^{\infty}_{\lambda=1}c_{\lambda}\phi_{\lambda}(\Phi_0)e^{-it\lambda}$ (as in Theorem 6), but now satisfying $i\partial_tU=-\Delta_2U$ and potential $V(x_1,x_2,x_3,t)=V(x^a,t)$, depended from $x^a$, $t$. Let also $G=G(x^a,t)$ be harmonic with respect to $x^a$, i.e. $\Delta_2G=0$. Then
$$
i\partial_t\Psi=-\Delta_2\Psi+V\Psi,
$$
or equivalently
$$
i\partial_t U+i\partial_t G=-\Delta_2U-\Delta_2G+VU+VG.
$$
Hence if also $U$ is such that $i\partial_tU=-\Delta_2 U$,  we have
$$
i\partial_t G=(U+G)V,
$$
or equivalently
$$
V=\frac{i\partial_tG}{U+G}.
$$
QED\\
\\
\textbf{Notes.}\\
1) All the potentials of the form
\begin{equation} 
V(x,y,z,t)=V(x,y,z)=\frac{c_1f\left(e^{-i \cot^{-1}\left(\frac{x}{y}\right)}\sqrt{x^2+y^2}\right)}{c_2f\left(e^{-i\cot^{-1}\left(\frac{x}{y}\right)}\sqrt{x^2+y^2}\right)-P\left(\sqrt{x^2+y^2+z^2}\right)},
\end{equation}
where $f$ is arbitrary smooth function of $\bf R\rm$ and $P(x)=\sum^{\infty}_{\lambda=1}c_{\lambda}\phi_{\lambda}(x)$, are solvable ($c_{\lambda}$ are arbitrary numbers such the series $P(x)$ converges).\\
\\
2) If $F$ is a smooth function and 
\begin{equation}
A(x_1,x_2,x_3)=\int\int\int_{\bf R^3\rm}\frac{F(x'_1,x'_2,x'_3)}{\left[1+\sum^{3}_{k=1}(x_k-x'_k)^2\right]^2}dx'_1dx'_2dx'_3,
\end{equation} 
then $A(x_1,x_2,x_3)$ is Harmonic function in $\bf R^3\rm$. The opposite is also true: For every harmonic function $A$ exists $F$ such that (64) holds.\\
\\
\textbf{Example.}\\
In the case of $\bf E_3\rm$, for $\partial V/\partial t=0$  and $0=c_2=c_3=\ldots$, we have that all the potentials 
$$
V(x_1,x_2,x_3)=\frac{A\left(x_1,x_2,x_3\right)}{2(x_1^2+x_2^2+x_3^2)^{-1/2}\cos\left(\sqrt{x_1^2+x_2^2+x_3^2}\right)-A\left(x_1,x_2,x_3\right)},
$$
are solvable.\\
\\

Theorem 11 can expanded to every metric non Eucledian space as follows\\
\\
\textbf{Theorem 12.}\\
Let  
$$
\overline{S}(x_1,x_2,\ldots,x_N)=
$$
$$
=\left\{S_1(x_1,x_2,\ldots,x_N),S_2(x_1,x_2,\ldots,x_N),\ldots,S_N(x_1,x_2,\ldots,x_N)\right\}
$$
be an arbitrary metric space. In this case we have $\nu=(N-2)/2$ and   
$$
\Delta_2\left(\phi_{\nu,\lambda}(\Phi_0)\right)=-\lambda\phi_{\nu,\lambda}(\Phi_0) ,\eqno{(64.1)} 
$$
where
$$
\phi_{\nu,\lambda}(x)=c_1x^{-\nu}J_{\nu}\left(\sqrt{\lambda}x\right)+c_2x^{-\nu}Y_{\nu}\left(\sqrt{\lambda}x\right).\eqno{(64.2)} 
$$
Also if $\Phi_0=\sqrt{\sum^{N}_{k=1}S_k(x_1,x_2,\ldots,x_N)^2}$,  $x^a=\left\{x_1,x_2,\ldots,x_N\right\}$, then for a given potential  
\begin{equation}
V(x^a,t)=\frac{\sum^{\infty}_{\lambda=1}c^{*}_{\lambda}\lambda e^{-it\lambda}A_{\lambda}(x^a)}{\sum^{\infty}_{\lambda=1}c^{*}_{\lambda} e^{-it\lambda}A_{\lambda}(x^a)+\sum^{\infty}_{\lambda=1}c_{\lambda}\phi_{\nu,\lambda}(\Phi_0)e^{-it\lambda}}, 
\end{equation}
with $\Delta_2(A_{\lambda}(x^a))=0$, a solution of
\begin{equation}
i\partial_t\Psi(x^a,t)=-\Delta_{2,x}\Psi(x^a,t)+V(x^a,t)\Psi(x^a,t),
\end{equation}     
is
\begin{equation}
\Psi(x^a,t)=\sum^{\infty}_{\lambda=1}c_{\lambda}e^{-it\lambda}\phi_{\nu,\lambda}(\Phi_0)+\sum^{\infty}_{\lambda=1}c^{*}_{\lambda}e^{-it\lambda}A_{\lambda}(x^a) .
\end{equation}
\\
\textbf{Theorem 13.}\\
If the potentials $V(x^a,t)$ are of the form
\begin{equation}
V(x^a,t)=\frac{\sum^{\infty}_{k=1}c^{*}_{k}k e^{-itk}A_{k}(x^a)}{\sum^{\infty}_{k=1}c^{*}_{k} e^{-it k}A_{k}(x^a)+\sum^{\infty}_{k=1}c_{k}\phi_{\nu,k}(\Phi_0)e^{-itk}} ,
\end{equation}
where 
$$
A_k(x^a)=\sum^{\infty}_{\lambda=1}a_{k,\lambda}e^{\sqrt{\lambda}S_N(x^a)}U_{\lambda}(\Phi^{-}_0),
$$
with $\Phi^{-}_0=\sqrt{\sum^{N-1}_{\lambda=1}S_k(x^a)^2}$, 
then a solution of (66) is 
\begin{equation}
\Psi=\Psi(x^{a},t)=\sum^{\infty}_{k=1}c_{k}e^{-itk}\phi_{\nu,k}(\Phi_0)+\sum^{\infty}_{k=1}c^{*}_{k}e^{-itk}A_{k}(x^a) .
\end{equation}
\\
\textbf{Proof.}\\
Observe that in any space $S$ holds
\begin{equation}
\Delta_2\left(A_k(x^a)\right)=0
\end{equation}
\\
\textbf{Theorem 14.}\\
The potentials of the form 
$$
V=V(x^a,t)=-\frac{\sum^{\infty}_{\lambda=1}c_{\lambda}\sqrt{\lambda}e^{-it\lambda}\phi_{\nu+1,\lambda}(\Phi_0)}{\sum^{\infty}_{\lambda=1}c_{\lambda}e^{-it\lambda}\phi_{\nu,\lambda}(\Phi_0)}=
$$
\begin{equation}
=\left(\frac{1}{x}\frac{\partial }{\partial x}\log\left(\sum^{\infty}_{\lambda=1}c_{\lambda}e^{-it\lambda}\phi_{\nu,\lambda}(x)\right)\right)_{x=\Phi_0},
\end{equation}
are solvable in view of Theorem 16 below.\\
\\
\textbf{Theorem 15.}\\
Consider the case of $S=\bf R\rm$. Then if
\begin{equation}
V(x,t)=-\frac{2\nu+1}{x}\partial_x\left(\log(u(x,t))\right)
\end{equation} 
and $u(x,t)$ satisfies the PDE
\begin{equation}
i\partial_t u(x,t)=-\partial^2_{x} u(x,t)-\frac{2\nu+1}{x}\partial_x u(x,t), 
\end{equation}
the Scrhodinger's equation
\begin{equation}
i\partial_t\Psi(x,t)=-\partial^2_{x}\Psi(x,t)+V(x,t)\Psi(x,t)
\end{equation}
is solvable and the solution is $\Psi(x,t)=u(x,t)$.\\
\\
\textbf{Theorem 16.}\\
The functions
\begin{equation}
u_0(x,t)=\sum^{\infty}_{\lambda=1}c_{\lambda}e^{-it\lambda}\phi_{\nu,\lambda}\left(x\right),
\end{equation}
are solutions of PDE
\begin{equation}
i\partial_t u(x,t)=-\partial^2_{x}u(x,t)-\frac{2\nu+1}{x}\partial_x u(x,t).
\end{equation}
The oposite is also true. If $u(x,t)$ satisfy (76), then $u(x,t)$ are of the form (75).\\ Also if the dimension of $S$ is $N$ and the potential's are of the form
\begin{equation}
V(x^a,t)=-\frac{2\nu+2-N}{\Phi_0}\left(\partial_x\log u(x,t)\right)_{x=\Phi_0},
\end{equation}
with $u(x,t)$ solution of (76), then equation 
\begin{equation}
i\partial_t\Psi(x^a,t)=-\Delta_{2,x}\Psi(x^a,t)+V(x^a,t)\Psi(x^a,t),
\end{equation}
have solution $\Psi(x^a,t)=u(\Phi_0,t)$.\\
\\

Assume now $r_{\nu,\lambda}$ is the $\lambda-$th root of $J_{\nu}(x)=0$ and $\rho_{\nu,\lambda}=r^2_{\nu,\lambda}$. When $f(x,t)\in D=L[0,1]\times \textbf{R}$, then the functions $x^{\nu+1/2}\phi_{\nu,\rho_{\nu,\lambda}}\left(x\right)$, (for $c_2=0$ in (64.2)) are bases in $D$ and writing 
\begin{equation}
f(x,t)=\sum^{\infty}_{p=1}C_{p}e^{-it\rho_{\nu,p}}x^{\nu+1/2}\phi_{\nu,\rho_{\nu,p}}(x),
\end{equation}
we have
$$
\int^{1}_{0}x^{2\nu+1}\phi_{\nu,\rho_{\nu,n}}\left(x\right)\phi_{\nu,\rho_{\nu,m}}\left(x\right)dx=\int^{1}_{0}xJ_{\nu}\left(xr_{\nu,n}\right)J_{\nu}\left(xr_{\nu,m}\right)dx=
$$
$$
=\frac{1}{2}J_{\nu+1}(r_{\nu,n})J_{\nu+1}(r_{\nu,m})\delta_{n,m}.
$$
The $\delta_{n,m}$ is 1 if $n=m$ and 0 otherwise. Hence the coefficients $C_{p}$ in (79) are
\begin{equation}
C_{p}=\frac{2e^{it\rho_{\nu,p}}}{J_{\nu+1}(r_{\nu,p})^2}\int^{1}_{0}f(x,t)x^{\nu+1/2}\phi_{\nu,\rho_{\nu,p}}\left(x\right)dx.
\end{equation}
In case 
$$
\frac{2e^{it\rho_{\nu,p}}}{J_{\nu+1}(r_{\nu,p})^2}\int^{1}_{0}f(x,t)x^{1/2}J_{\nu}\left(r_{\nu,p}x\right)dx
$$
is independed of $t$ we have that indeed $f(x,t)$ is solution of (76). Hence 
\begin{equation}
\int^{1}_{0}f(x,t)x^{\nu+1/2}\phi_{\nu,\rho_{\nu,p}}\left(x\right)dx=\frac{1}{2}C_pJ_{\nu+1}\left(r_{\nu,p}\right)^2e^{-itr_{\nu,p}^2},
\end{equation}
where $C_p$ is constant.\\ Also assuming the initial condition $f(x,0)=f_0(x)$, we get
\begin{equation}
f(x,t)=\sum^{\infty}_{p=1}\left(\frac{2x^{-\nu}}{J_{\nu+1}(r_{\nu,p})^2}\int^{1}_{0}f_0(y)y^{1/2}J_{\nu}\left(r_{\nu,p}y\right)dy\right)e^{-ir_{\nu,p}^2t}J_{\nu}\left(r_{\nu,p}x\right).
\end{equation}
Now if we assume that the function $f(x,t)$ have an expansion of the form
\begin{equation}
f(x,t)=\sum^{\infty}_{q=1}c_qA_{q}(t)B_q(x),
\end{equation}
then
$$
\int^{1}_{0}f(x,t)x^{\nu+1/2}\phi_{\nu,\rho_{\nu,p}}\left(x\right)dx=\sum^{\infty}_{q=1}c_qA_q(t)\int^{1}_{0}B_q(x)x^{1/2}J_{\nu}(r_{\nu,p}x)dx=
$$
$$
=\frac{1}{2}C_pJ_{\nu+1}(r_{\nu,p})^2e^{-itr_{\nu,p}^2}.
$$
Hence there exists constants $C_{\nu,q,p,}$, $E_{\nu,p}$, such that
\begin{equation}
\sum^{\infty}_{q=1}C_{\nu,q,p}A_q(t)=E_{\nu,p}e^{-itr_{\nu,p}^2}\textrm{, }\forall t\in\textbf{R}.
\end{equation}
Write
\begin{equation}
A_q(t)=\sum^{\infty}_{l=1}\eta_{q,l}e^{-itr_{\nu,l}^2}.
\end{equation}
Then
$$
\sum^{\infty}_{q=1}C_{\nu,q,p}\sum^{\infty}_{l=1}\eta_{q,l}e^{-itr_{\nu,l}^2}=E_{\nu,p}e^{-itr_{\nu,p}^2}\Rightarrow
$$
$$
\sum^{\infty}_{l=1}\left(\sum^{\infty}_{q=1}\eta_{q,l}C_{\nu,q,p}\right)e^{-itr_{\nu,l}^2}=E_{\nu,p}e^{-itr_{\nu,p}^2}\Rightarrow
$$
\begin{equation}
\sum^{\infty}_{q=1}\eta_{q,l}C_{\nu,q,p}=\delta_{p,l}E_{\nu,l}.
\end{equation}
Hence also
\begin{equation}
\sum^{\infty}_{q=1}\eta_{q,p}C_{\nu,q,p}=E_{\nu,p}.
\end{equation}
Assume that $\eta^{*}_{l,a}$ is such that $\sum^{\infty}_{l=1}\eta_{q,l}\eta^{*}_{l,a}=\delta_{q,a}$, then
$$
\sum^{\infty}_{q=1}\sum^{\infty}_{l=1}\eta_{q,l}\eta^{*}_{l,a}C_{\nu,q,p}=\sum^{\infty}_{l=1}\delta_{p,l}\eta^{*}_{l,a}E_{\nu,l}\Rightarrow\sum^{\infty}_{q=1}\delta_{q,a}C_{\nu,q,p}=\eta^{*}_{p,a}E_{\nu,p}\Rightarrow
$$
\begin{equation}
C_{\nu,a,p}=\eta^{*}_{p,a}E_{\nu,p}.
\end{equation}
Hence $A_q(t)$ have the form (85) and staisfy (86),(87). From (86) with $p\neq l$, we have 
$$
\sum^{\infty}_{q=1}\eta_{q,l}C_{\nu,q,p}=0\textrm{, }l\neq p.
$$
Hence
$$
\sum^{\infty}_{q=1}\sum_{l\neq p}\eta_{q,l}\eta^{*}_{l,a}C_{\nu,q,p}=0\Leftrightarrow
$$
$$
\sum^{\infty}_{q=1}\left[\left(\sum^{\infty}_{l=1}\eta_{q,l}\eta^{*}_{l,a}\right)-\eta_{q,p}\eta^{*}_{p,a}\right]C_{\nu,q,p}=0\Leftrightarrow
$$
$$
\sum^{\infty}_{q=1}\left(\delta_{q,a}-\eta_{q,p}\eta^{*}_{p,a}\right)C_{\nu,q,p}=0\Leftrightarrow
$$
$$
C_{\nu,a,p}-\eta^{*}_{p,a}\sum^{\infty}_{q=1}\eta_{q,p}C_{\nu,q,p}=0\Leftrightarrow
$$
$$
\eta^{*}_{p,a}E_{\nu,p}-\eta^{*}_{p,a}\sum^{\infty}_{q=1}\eta_{q,p}C_{\nu,q,p}=0.
$$
Hence $\eta^{*}_{p,a}=0$ or if $\eta^{*}_{p,a}\neq 0$, we have
$$
E_{\nu,p}=\sum^{\infty}_{q=1}\eta_{q,p}C_{\nu,q,p}.
$$
Now since $x^{\nu+1/2}\phi_{\nu,\rho_{\nu,l}}=x^{1/2}J_{\nu}(r_{\nu,l}x)$ is a base of $L[0,1]$ every function $B_{q}(x)$ have expansion of the form
\begin{equation}
B_{q}(x)=B_{\nu,q}(x)=\sum^{\infty}_{k=1}c^{*}_{\nu,q,k}x^{\nu+1/2}\phi_{\nu,\rho_{\nu,k}}(x)dx=x^{1/2}\sum^{\infty}_{k=1}c^{*}_{\nu,q,k}J_{\nu}\left(r_{\nu,k}x\right).
\end{equation}
Hence
$$
c_q\int^{1}_{0}B_{q}(x)x^{\nu+1/2}\phi_{\nu,\rho_{\nu,q}}\left(x\right)dx=C_{\nu,q,p}\Leftrightarrow
$$
$$
c_qc^{*}_{\nu,q,p}\frac{1}{2}J_{\nu+1}(r_{\nu,p})^2=C_{\nu,q,p}.
$$
Using (89) and (85) in (83) we get
$$
f(x,t)=\sum^{\infty}_{q=1}c_q\left(\sum^{\infty}_{l=1}\eta_{q,l}e^{-itr^2_{\nu,l}}\right)\left(x^{1/2}\sum^{\infty}_{k=1}c^{*}_{\nu,q,k}J_{\nu}\left(r_{\nu,k}x\right)\right)=
$$
$$
=x^{1/2}\sum^{\infty}_{q,l,k=1}c_q\eta_{q,l}c^{*}_{\nu,q,k}e^{-itr^2_{\nu,l}}J_{\nu}\left(r_{\nu,k}x\right)=
$$
$$
=\frac{2x^{1/2}}{J_{\nu+1}(r_{\nu,p})^2}\sum^{\infty}_{q,k,l=1}\eta_{q,l}C_{\nu,q,k}e^{-itr^2_{\nu,l}}J_{\nu}\left(r_{\nu,k}x\right)=
$$
$$
=\frac{2x^{1/2}}{J_{\nu+1}(r_{\nu,p})^2}\sum^{\infty}_{k,l=1}\delta_{k,l}E_{\nu,l}e^{-itr^2_{\nu,l}}J_{\nu}\left(r_{\nu,k}x\right)=
$$
$$
=\frac{2x^{1/2}}{J_{\nu+1}(r_{\nu,p})^2}\sum^{\infty}_{k=1}E_{\nu,k}e^{-itr^2_{\nu,k}}J_{\nu}\left(r_{\nu,k}x\right)=
$$
$$
=\frac{2x^{1/2}}{J_{\nu+1}(r_{\nu,p})^2}\sum^{\infty}_{k=1}\frac{1}{2}C_kJ_{\nu+1}(r_{\nu,k})^2e^{-itr^2_{\nu,k}}J_{\nu}\left(r_{\nu,k}x\right)=
$$
$$
=x^{1/2}\sum^{\infty}_{k=1}C_ke^{-itr^2_{\nu,k}}J_{\nu}\left(r_{\nu,k}x\right).
$$
Hence we have the next\\
\\
\textbf{Theorem 17.}\\
Assume that given $f(x,t)$, this is of the form (for example from a PDE problem):
\begin{equation}
f(x,t)=\sum^{\infty}_{p=1}c_pA_{p}(t)B_p(x).
\end{equation} 
Assume also that 
\begin{equation}
e^{itr_{\nu,p}^2}\int^{1}_{0}f(x,t)x^{1/2}J_{\nu}\left(r_{\nu,p}x\right)dx=\frac{1}{2}C_pJ_{\nu+1}\left(r_{\nu,p}\right)^2,
\end{equation}
for some constants $C_p$. Then $f(x,t)$ satisfies the PDE
\begin{equation}
i\partial_t u(x,t)=-\partial^2_{x}u(x,t)-\frac{2\nu+1}{x}\partial_x u(x,t)
\end{equation}
and the general solution $u(x,t)$ of (92) as also the function $f(x,t)$ are given by
\begin{equation}
u(x,t)=f(x,t)=x^{1/2}\sum^{\infty}_{p=1}C_pe^{-itr_{\nu,p}^2}J_{\nu}(r_{\nu,p}x).
\end{equation}

\section{The self-adjoint property of the Hamiltonian}

The Hamiltonian of equation (11) in the space 
\begin{equation}
S : \overline{\bf S\rm}(x^a)=\{S_1(x^a),S_2(x^a),\dots,S_N(x^a)\},
\end{equation} 
is
\begin{equation}
H\Psi=-\Delta_{2,x}\Psi+V(x^a,t)\Psi.
\end{equation}    
We define the inner product of the functions $f=f(x^a)$ and $g=g(x^a)$ with
\begin{equation}
\left\langle f,g\right\rangle:=\int_{\bf R^N\rm}f(x_1,x_2,\ldots,x_N)\overline{g(x_1,x_2,\ldots,x_N)}dx_1dx_2\ldots dx_N=\int_{\bf R^N\rm}f\overline{g}dx^a.
\end{equation}
We will find the conditions under in what spaces and what functions $F_1$, $G_1$ holds.  
\begin{equation}
\int_{\bf R^N\rm}\left(HF_1(\Phi_0)\overline{G_1(\Phi_0)}-\overline{HG_1(\Phi_0)}F_1(\Phi_0)\right)dx^a=0.
\end{equation}
We assume first that $F_1$, $G_1$ are single valued. We have 
\begin{equation}
\int_{\bf R^N\rm}\left[\overline{G_1(\Phi_0)}\Delta_2F_1(\Phi_0)-F_1(\Phi_0)\overline{\Delta_2G_1(\Phi_0)}\right]dx^a=0.
\end{equation} 
For this we make change of variables of the form 
$$
x'_i=f_i(x^a) \textrm{ , } i=1,2,\ldots,N.
$$
More precisely we set
$$  
x'_k=S_k(x^a) \textrm{ , } k=1,2,\ldots,N. \eqno{(T)} 
$$ 
Then equation (98) is transformed into
$$
\int_{\bf R^N\rm}[\overline{G_1\left(\sqrt{x'^2_1+x'^2_2+\ldots+x'^2_N}\right)}\Delta_2F_1\left(\sqrt{x'^2_1+x'^2_2+\ldots+x'^2_N}\right)-
$$
$$
-F_1\left(\sqrt{x'^2_1+x'^2_2+\ldots+x'^2_N}\right)\overline{\Delta_2G_1\left(\sqrt{x'^2_1+x'^2_2+\ldots+x'^2_N}\right)}]\times 
$$
$$
\times\frac{D(x_1,x_2,\ldots,x_N)}{D(x'_1,x'_2,\ldots,x'_N)}dx'_1dx'_2\ldots dx'_N=0. 
$$
Applying again the change of variables into polar-type $$x'_1=r\cos(\theta_1)\cos(\theta_2)\ldots\cos(\theta_{N-2})\cos(\theta_{N-1}),$$
$$x'_2=r\cos(\theta_1)\cos(\theta_2)\ldots\cos(\theta_{N-2})\sin(\theta_{N-1}),$$
$$
x'_3=r\cos(\theta_1)\cos(\theta_2)\ldots\cos(\theta_{N-3})\sin(\theta_{N-2}),
$$
$$x'_4=r\cos(\theta_1)\cos(\theta_2)\ldots\cos(\theta_{N-4})\sin(\theta_{N-3}),$$
$$\vdots$$
$$x'_{N-1}=r\cos(\theta_1)\sin(\theta_{2}),$$
\begin{equation}
x'_N=r\sin(\theta_{1}),
\end{equation}
and 
$A_{pol}=\{0\leq r<+\infty$, $\theta_1,\theta_2,\ldots,\theta_{N-2}\in\left(-\pi/2,\pi/2\right)$, $\theta_{N-1}\in\left(0,2\pi\right)\}$,\\  
we have to show:
$$
\int_{A_{pol}}[G_1\left(r\right)\left(\frac{N-1}{r}F_1^{(1)}(r)+F_1^{(2)}(r)\right)
-F_1\left(r\right)\left(\frac{N-1}{r}G_1^{(1)}(r)+G_1^{(2)}(r)\right)]\times
$$
$$
\times\frac{D(x_1,x_2,\ldots,x_N)}{D(x'_1,x'_2,\ldots,x'_N)}\frac{D(x'_1,x'_2,\ldots,x'_N)}{D(r,\theta_2,\ldots,\theta_{N-1})}dr d\theta_1\ldots d\theta_{N-1}=0. 
$$    
Or if we assume that 
\begin{equation}
\frac{D(x_1,x_2,\ldots,x_N)}{D(x'_1,x'_2,\ldots,x'_N)}\frac{D(x'_1,x'_2,\ldots,x'_N)}{D(r,\theta_2,\ldots,\theta_{N-1})}
=r^{(N-1)\mu}T(\theta_1,\theta_2,\ldots,\theta_{N-1}).
\end{equation} 
Then equivalently we have 
$$
\int^{\infty}_{0}r^{(N-1)\mu}[G_1(r)\left(\frac{N-1}{r}F_1'(r)+F_1''(r)\right)
-F_1(r)\left(\frac{N-1}{r}G_1'(r)+G_1''(r)\right)]dr\times
$$
$$
\times \int T(\theta_1,\theta_2,\ldots,\theta_{N-1})\cos(\theta_1)^{N-2}\cos(\theta_2)^{N-1}\ldots \cos(\theta_{N-1}) d\theta_1\ldots d\theta_{N-1}=0,
$$ 
which is true for functions in $\textbf{F}^{N}_{\mu}$, since the following integral expansion is valid  
$$
\int^{\infty}_{0} r^{(N-1)\mu}\left(G_1(r)\left(\frac{N-1}{r}F_1'(r)+F_1''(r)\right)
-F_1(r)\left(\frac{N-1}{r}G_1'(r)+G_1''(r)\right)\right)dr=
$$
$$
=\left[r^{(N-1)\mu}\left(G_1(r)F'_1(r)-F_1(r)G'_1(r)\right)\right]^{\infty}_{0}-
$$
$$
-(\mu-1)(N-1)\int^{\infty}_{0}r^{(N-1)\mu-1}\left(F'(r)G_1(r)-F_1(r)G_1'(r)\right)dr=0.
$$

Hence we conclude that\\
\\
\textbf{Theorem 18.}\\If in the metric space $S$ the transformation ($T$) have the property 
\begin{equation}
\frac{D(x_1,x_2,\ldots,x_N)}{D(x'_1,x'_2,\ldots,x'_N)}\frac{D(x'_1,x'_2,\ldots,x'_N)}{D(r,\theta_2,\ldots,\theta_{(N-1)\mu})}
=r^{(N-1)\mu}T(\theta_1,\theta_2,\ldots,\theta_{N-1}),
\end{equation}  
then the Hamiltonian is a self adjoint operator over $\textbf{F}^{N}_{\mu}$ (Definition 4, pg 3), provided that $T(\theta_1,\theta_2,\ldots,\theta_{N-1})$ is continuous.\\
\\
A consequence of this is the next\\
\\
\textbf{Theorem 19.}\\
In a space $S^{N}_{\mu}$ the Hamiltonian $H$ is self adjoint over $\textbf{F}^{N}_{\mu}$.\\
\\
\textbf{Proof.}\\
If the transformation is $x'_k(x^a)=S_k(x^a)$, $k=1,2,\ldots,N$ and is homogeneous of degree $\mu$ then the inverse is $x_k=S'_k\left(x'^a\right)$ and is also homogeneous of degree $\mu$. 
Also $\Phi_0(x^a)=\sqrt{\sum^{N}_{k=1}S_k(x^a)^2}$ is homogeneous of degree $\mu$ and every first order partial derivative is of degree $\mu-1$. Hence the determinant 
$$
\frac{D(x_1,x_2,\ldots,x_N)}{D(x'_1,x'_2,\ldots,x'_N)}=\textrm{det}\left(H_{k,j}\right)=\textrm{det}\left[\left(\frac{\partial S'_k\left(x'_1,x'_2,\ldots,x'_N\right)}{\partial x'_j}\right)_{k,j}\right],
$$ 
is homogeneous function of degree $(\mu-1)N$.  
$$
dx_1dx_2\ldots dx_N=\frac{D(x_1,x_2,\ldots,x_N)}{D(x'_1,x'_2,\ldots,x'_N)}dx'_1dx'_2\ldots dx'_N=
$$
$$
=E(x'_1,x'_2,\ldots,x'_N)dx'_1dx'_2\ldots dx'_N=
$$
$$
=\textrm{det}\left[\left(\frac{\partial S'_k\left(x'_1,x'_2,\ldots,x'_N\right)}{\partial x'_j}\right)_{k,j}\right]dx'_1dx'_2\ldots dx'_N.
$$
Setting now $x'_j\rightarrow \xi y'_j$, $j=1,2,\ldots,N$: ($\sigma$), we have 
$$dx'_1dx'_2\ldots dx'_N=\xi^{N}dy'_1dy'_2\ldots dy'_N.$$ Hence $E(x'_1,x'_2,\ldots,x'_N)$ is homogeneous of degree $\mu N$. Lastly, taking the polar coordinates (99), from the homogenicity of $E$ we have that $r$ is a common factor (i.e $r$ behaves like $\xi$) and for a certain $T$ holds  
$$
E=r^{(N-1)\mu}T(\theta_1,\theta_2,\ldots,\theta_{N-1})\textrm{ , }\partial_r T=0.
$$
This is condition (101) of Theorem 18.

\section{The Eucledian Space and the Polar Coordinates}

In the Euclidean space $\textbf{E}_3$ the metric is $g_{ij}=\delta_{ij}$ and
$$
\Delta_2(\Phi)=\frac{\partial^2 \Phi}{\partial x^2}+\frac{\partial^2 \Phi}{\partial y^2}+\frac{\partial^2 \Phi}{\partial z^2}.
$$
We make the change of coordinates
$$
x=r\sin(\theta)\cos(\phi) , y=r\sin(\theta)\sin(\phi) , z=r\cos(\theta).
$$
Then 
$$
\Delta_2(\Phi)=\frac{\partial^2 \Phi}{\partial r^2}+\frac{2}{r}\frac{\partial \Phi}{\partial r}+\frac{1}{r^2}\left(\frac{1}{\sin^2(\theta)}\frac{\partial^2\Phi}{\partial \phi^2}+\cot(\theta)\frac{\partial \Phi}{\partial \theta}+\frac{\partial^2 \Phi}{\partial \theta^2}\right)
$$   
and holds
\begin{equation}
\Delta_2\left(f(e^{-i\phi}r\sin(\theta))\right)=\Delta_1\left(f(e^{-i\phi}r\sin(\theta))\right)=0.
\end{equation}
\\

Expanding this idea in $\textbf{E}_N$-Eucledian space with dimension $N=2n$, for any function $f$ analytic at the origin we have\\
\\
\textbf{Theorem 20.}\\
For $f$ analytic in the origin set
\begin{equation}
\Phi_0(r,\theta_1,\ldots,\theta_{N-1})=re^{i\theta_1}e^{\psi(\theta_2)}e^{i\theta_3}e^{\psi(\theta_4)}\ldots e^{i\psi(\theta_{N-2})}e^{i\theta_{N-1}},
\end{equation} 
where $N=2n$, $n=1,2,3,\ldots$ and  
\begin{equation}
\psi(x)=2\textrm{arctanh}\left(\tan\left(\frac{x}{2}\right)\right). 
\end{equation}
Then
\begin{equation}
\Delta_{1S}(f(\Phi_0))=0.
\end{equation}
For every $f$.\\
\\ 
\textbf{Proof.}\\
If $x=(x_1,x_2,\ldots,x_N)$ is an element of $\textbf{R}^N$ then we have 
$$
r=\sqrt{x^2_1+x^2_2+\ldots+x^2_N}
$$
and 
$$
\theta_1,\theta_2,\ldots,\theta_{N-1}\in \left(-\frac{\pi}{2},\frac{\pi}{2}\right),
$$
$$
\theta_{N-1}\in(0,2\pi),
$$
then
$$x_1=r\cos(\theta_1)\cos(\theta_2)\ldots\cos(\theta_{N-2})\cos(\theta_{N-1}),$$
$$x_2=r\cos(\theta_1)\cos(\theta_2)\ldots\cos(\theta_{N-2})\sin(\theta_{N-1}),$$
$$
x_3=r\cos(\theta_1)\cos(\theta_2)\ldots\cos(\theta_{N-3})\sin(\theta_{N-2}),
$$
$$x_4=r\cos(\theta_1)\cos(\theta_2)\ldots\cos(\theta_{N-4})\sin(\theta_{N-3}),$$
$$\vdots$$
$$x_{N-1}=r\cos(\theta_1)\sin(\theta_{2}),$$
\begin{equation}
x_N=r\sin(\theta_{1}).
\end{equation}
Writing 
$$
\Phi_1(u_1,u_2,\ldots,u_N)=a_1(u_1)a_2(u_2)\ldots a_N(u_N),
$$
we try to solve the spherical partial differential equation
$$\Delta_{1S}(\Phi_1)=0$$
(by $\Delta_{1S}$ we note the $1-$Beltrami operator in spherical coordinates).\\
One can mange to find $g_{ij}$ at any rate $N$, using  Mathematica Program. We first find the solutions for $N=2,3,4,5,6,7$ and then we derive numerically our results for arbitrary dimensions. Actually the operator $\Delta_{1S}(\Phi_1)$ is quite simple, since the $g_{ij}$ form a diagonal matrix. We have:
$$
\left(u_1\prod^{N}_{k=1}a_k(u_k) \right)^{-2}\Delta_{1S}(\Phi_1)=
$$
\begin{equation}
=\frac{a'_1(u_1)^2u_1^2}{a_1(u_1)^2}+\frac{a'_2(u_2)^2}{a_2(u_2)^2}+\sum^{N}_{k=3}\frac{a'_k(u_k)^2}{a_k(u_k)^2}\prod^{k-1}_{j=2}\sec(u_{j})^2=0.
\end{equation} 
Hence the problem reduces to prove that $g_{ij}$ is diagonal and the values in the diagonal are
$$ 
\left\{1,u_1^2,u_1^2\cos(u_2)^2,u_1^2\cos(u_2)^2\cos(u_3)^2,\ldots,u_1^2\prod^{N-1}_{j=2}\cos(u_j)^2\right\}.
$$
If we assure that, then we solve the three simple differential equations $xy'/y=c$, $y'/y=1$ and $y'/y=c\sec(x)$ and the result follows.\\
\\

Another interesting proposition for evaluations of spherical $1-$Beltrami operators is\\
\\
\textbf{Theorem 21.}\\
Let $f$ be analytic in a open set containing the origin, then
if $N=2n$, $n=2,3,\ldots$ and
$$ G(x_1,x_2,x_3,x_4,\ldots,x_{N-3},x_{N-2},x_{N-1},x_N)=
$$
$$
=f_1(x_1x_2)f_2(x_3x_4)f(x_5x_6)\ldots f_{n-1}(x_{N-3}x_{N-2})f_n(x_{N-1}x_N),
$$
where $x_1=r$, $x_2=e^{i\theta_1}$, $x_3=e^{\psi(\theta_2)}$, $x_4=e^{i \theta_3}$, $x_5=e^{\psi(\theta_4)}$, $\ldots$, $x_{N-1}=e^{\psi(\theta_{N-2})}$, $x_N=e^{i\theta_{N-1}}$, then also
\begin{equation}
\Delta_{1S}(G)=0,
\end{equation}
where $f_j$, $j=1,2,\ldots,n$ are single valued.\\
\\ 
\textbf{Proof.}\\
Equation (108) follows from (107) where we can set the terms of (107) respectively with $p_1,p_2,\ldots,p_{N-1},p_N$ and then take $p_k+p_{k+1}=0$, for the equation to hold (Note that we use, as in Theorem 20, the method of separate variables). The $p$'s in the general solution appear as powers $e^{i(p_k\theta_k+p_{k+1}\psi(\theta_{k+1}))}$.

\section{Appendix. (The general solution of space independed Schrodinger equation)}

I have read many books in Quantum physics that treat with the subject of time independent Schrodinger equation (TISE). None of these books have treated with SISE (space independed Schrodinger equation). I thought that would be nice to present a general solution of SISE here.\\
\\

The SISE in a potential $V(t)$ read as
$$i\hbar\frac{\partial}{\partial t} \Psi(x,t)=-\frac{\hbar^2}{2m}\frac{\partial^2}{\partial x^2} \Psi(x,t)+V(t)\Psi(x,t).\eqno{(1)}$$

For a smooth function $\Psi(x,t)$ in a certain $L_p$ space, the Fourier transform is given from  
$$
\widehat{\Psi}(s,t)=\int_{\textbf{\scriptsize R\normalsize}}\Psi(x,t)e^{-ixs}dx\Leftrightarrow\Psi(x,t)=\frac{1}{2\pi}\int_{\textbf{\scriptsize R\normalsize}}\widehat{\Psi}(s,t)e^{isx}ds.\eqno{(2)}
$$
Assume also that $\Psi$ satisfies the conditions 
$$
\lim_{x\rightarrow\pm\infty}\Psi(x,t)=\lim_{x\rightarrow\pm\infty}\partial_x\Psi(x,t)=0.\eqno{(3)}
$$
Taking the Fourier transform in (1) with respect to $x$, we have
$$
i\hbar\frac{\partial}{\partial t} \int_{\textbf{\scriptsize R\normalsize}}\Psi(x,t)e^{-ixs}dx=-\frac{\hbar^2}{2m}\int_{\textbf{\scriptsize R\normalsize}}\frac{\partial^2}{\partial x^2} \Psi(x,t)e^{-ixs}dx+V(t)\widehat{\Psi}(s,t)\Leftrightarrow
$$
$$
i\hbar\partial_t\widehat{\Psi}(s,t)=-\frac{\hbar^2}{2m}\left(\left[\partial_x\Psi(x,t)e^{-ixs}\right]^{x=+\infty}_{x=-\infty}+is\int_{\textbf{\scriptsize R\normalsize}}\frac{\partial}{\partial x} \Psi(x,t)e^{-ixs}dx\right)+
$$
$$
+V(t)\widehat{\Psi}(s,t)\Leftrightarrow
$$
$$
i\hbar\partial_t\widehat{\Psi}(s,t)=-\frac{\hbar^2}{2m}is\int_{R}\frac{\partial}{\partial x} \Psi(x,t)e^{-ixs}dx+V(t)\widehat{\Psi}(s,t)\Leftrightarrow
$$
$$
i\hbar\partial_t\widehat{\Psi}(s,t)=-\frac{\hbar^2}{2m}is\left(\left[\Psi(x,t)e^{-ixs}\right]^{x=+\infty}_{x=-\infty}+is\int_{\textbf{\scriptsize R\normalsize}} \Psi(x,t)e^{-ixs}dx\right)+
$$
$$
+V(t)\widehat{\Psi}(s,t)\Leftrightarrow
$$
$$
i\hbar\partial_t\widehat{\Psi}(s,t)=\left(\frac{\hbar^2}{2m}s^2+V(t)\right)\widehat{\Psi}(s,t)\eqno{(4)}
$$
This last equation can be solved easily. We get
$$
i\hbar\log\left(\widehat{\Psi}(s,t)\right)=\frac{\hbar^2}{2m}s^2t+\int V(t)dt+C(s)\Leftrightarrow
$$
$$
\widehat{\Psi}(s,t)=\exp\left(-\frac{i}{\hbar}C(s)\right)\exp\left(-\frac{i\hbar}{2m}s^2t-\frac{i}{\hbar}\int V(t)dt\right).
$$
Lastly inverting the Fourier transform we get
$$
\Psi(x,t)=\frac{1}{2\pi}\int_{\textbf{\scriptsize R\normalsize}}\exp\left(-\frac{i}{\hbar}C(s)\right)\exp\left(-\frac{i\hbar}{2m}s^2t-\frac{i}{\hbar}\int V(t)dt\right)e^{isx}ds
$$
and the general solution of (1) is
$$
\Psi(x,t)=\frac{1}{2\pi}\exp\left(-\frac{i}{\hbar}\int V(t)dt\right)\int_{\textbf{\scriptsize R\normalsize}}\exp\left(-\frac{i}{\hbar}C(s)-\frac{i\hbar}{2m}s^2t\right)e^{isx}ds,\eqno{(5)}
$$
where $C(s)$ is arbitrary smooth function defined from the needs of the problem. Actualy then
$$
\left|\widehat{\Psi}(s,t)\right|=\exp\left(\frac{Im(C(s))}{\hbar}\right).\eqno{(6)}
$$
Hence from Parseval's Identity: 
$$
f,g\in L_2\left(\textbf{R}\right)\Rightarrow\int_{\textbf{\scriptsize R\normalsize}}f(t)\overline{g(t)}dt=\frac{1}{2\pi}\int_{\textbf{\scriptsize R \normalsize}}\widehat{f}(\gamma)\overline{\widehat{g}(\gamma)}d\gamma,
$$
we get
$$
\int_{\textbf{\scriptsize R\normalsize}}\left|\Psi(x,t)\right|^2dx=\frac{1}{2\pi}\int_{\textbf{\scriptsize R\normalsize}}\left|\widehat{\Psi}(s,t)\right|^2ds=
\frac{1}{2\pi}\int_{\textbf{\scriptsize R\normalsize}}\exp\left(\frac{2Im(C(s))}{\hbar}\right)ds=
$$
$$
=c_0=constant.
$$
Hence the normalized solution is
$$
\Psi(x,t)=\frac{1}{2\pi\sqrt{c_0}}\exp\left(-\frac{i}{\hbar}\int V(t)dt\right)\int_{\textbf{\scriptsize R\normalsize}}\exp\left(-\frac{i}{\hbar}C(s)-\frac{i\hbar}{2m}s^2t\right)e^{isx}ds,\eqno{(7)}
$$
where
$$
c_0=\frac{1}{2\pi}\int_{\textbf{\scriptsize R\normalsize}}\exp\left(\frac{2Im(C(s))}{\hbar}\right)ds.\eqno{(8)}
$$
Hence the probability density is
$$
P(x,t)=\left|\Psi(x,t)\right|^2=\frac{1}{4\pi^2 c_0}\left|\int_{\textbf{\scriptsize R\normalsize}}\exp\left(-\frac{i}{\hbar}C(s)-\frac{i\hbar}{2m}s^2t\right)e^{isx}ds\right|^2
$$
and is indepentent from the potential. If we define as usual the inner product
$$
\left\langle f,g\right\rangle=\int^{+\infty}_{-\infty}f(t)\overline{g(t)}dt\textrm{, }\forall f,g\in L_2(\textbf{R})
$$
and the norm of $L_2(\textbf{R})$:
$$
||f||_2:=\sqrt{\left\langle f,f\right\rangle}=\sqrt{\int^{+\infty}_{-\infty}|f(t)|^2dt}\textrm{, }\forall f\in L_2(\textbf{R}).
$$
The expected value of a operator $\textbf{A}$ is
$$
\left\langle \textbf{A}\right\rangle=\left\langle \textbf{A} \Psi(x,t),\Psi(x,t)\right\rangle.
$$
One can easily see that the operator of energy is $\textbf{E}=i\hbar \partial_t$ and the operator of momentum $\textbf{p}=-i\hbar \partial_x$ and both are self-adjoint. The Hamiltonian of the system is
$$
\textbf{H}=\frac{\textbf{p}^2}{2m}+V(t).
$$
Also holds
$$
\textbf{E}\Psi=\textbf{H}\Psi
$$  
and this is (1). We will evaluate the expected value of energy $\textbf{E}$. 
$$
\left\langle \textbf{E}\right\rangle=\left\langle \textbf{E}\Psi, \Psi\right\rangle=\frac{1}{2\pi}\left\langle i\hbar \partial_t \widehat{\Psi},\widehat{\Psi}\right\rangle=\frac{1}{2\pi}\left\langle\left(\frac{\hbar^2}{2m}s^2+V(t)\right)\widehat{\Psi},\widehat{\Psi}\right\rangle\Rightarrow
$$
$$
\left\langle \textbf{E}\right\rangle=\frac{\hbar^2}{4\pi m}\int^{+\infty}_{-\infty}s^2|\widehat{\Psi}|^2ds+V(t).\eqno{(8.1)}
$$
Hence exists constant $c_1$ such that
$$
\left\langle \textbf{E}\right\rangle=V(t)+c_1.
$$
From this we get that energy is not conserved. However the expected value of the momentum is
$$
\left\langle \textbf{p}\right\rangle=\left\langle -i\hbar\partial_x \Psi,\Psi\right\rangle=-i\hbar\left\langle \partial_x \Psi,\Psi\right\rangle=\frac{-i\hbar}{2\pi}\left\langle \int_{\textbf{\scriptsize R\normalsize}}\partial_x\Psi(x,t) e^{-i x s}dx,\widehat{\Psi}(s,t)\right\rangle=
$$
$$
=\frac{-i\hbar}{2\pi}\left\langle \int_{\textbf{\scriptsize R\normalsize}}\partial_x\Psi(x,t) e^{-i x s}dx,\widehat{\Psi}(s,t)\right\rangle=
$$
$$
=\frac{-i \hbar}{2\pi}\left\langle-\int_{\textbf{\scriptsize R\normalsize}}\Psi(x,t)(-is)e^{-ixs}dx,\widehat{\Psi}(s,t)\right\rangle=
$$
$$
=\frac{\hbar}{2\pi}\left\langle s\widehat{\Psi},\widehat{\Psi}\right\rangle=\frac{\hbar}{2\pi}\int_{\textbf{\scriptsize R\normalsize}}s|\widehat{\Psi}(s,t)|^2ds\Rightarrow
$$
$$
\left\langle \textbf{p}\right\rangle=\frac{\hbar}{2\pi}\int_{\textbf{\scriptsize R\normalsize}}s\exp\left(\frac{2Im(C(s))}{\hbar}\right)ds=\frac{\hbar}{2\pi}\int_{\textbf{\scriptsize R\normalsize}}s|\widehat{\Psi}|^2ds=constant.\eqno{(8.2)}
$$
Since the expected value of momentum is constant, the momentum is conserved. We also can write
$$
\left\langle \textbf{p}^2\right\rangle=\left\langle -\hbar^2\partial^2_x\Psi,\Psi\right\rangle=\left\langle 2mi\hbar\partial_t\Psi-2mV(t)\Psi,\Psi\right\rangle=
$$
$$
=2m\left\langle \textbf{E}\right\rangle-2mV(t)=2m V(t)+2mc_1-2mV(t)=2mc_1=
$$
$$
=\frac{\hbar^2}{2\pi}\int^{+\infty}_{-\infty}s^2|\widehat{\Psi}|^2ds.\eqno{(8.3)}
$$
Hence
$$
\left(\Delta\textbf{p}\right)^2=\left\langle \textbf{p}^2\right\rangle-\left\langle \textbf{p}\right\rangle^2=\frac{\hbar^2}{2\pi}\int^{+\infty}_{-\infty}s^2|\widehat{\Psi}|^2ds-\frac{\hbar^2}{4\pi^2}\left(\int_{\textbf{\scriptsize R\normalsize}}s|\widehat{\Psi}|^2ds\right)^2=const.\eqno{(8.4)}
$$
The mean value of position is
$$
\left\langle \textbf{x}\right\rangle=\left\langle x\Psi,\Psi\right\rangle=\frac{1}{2\pi}\left\langle \left(\widehat{x\Psi}\right),\widehat{\Psi}\right\rangle=\frac{1}{2\pi}\left\langle \int^{+\infty}_{-\infty}x\Psi(x,t)e^{-i x s}dx,\widehat{\Psi}\right\rangle=
$$
$$
=\frac{i}{2\pi}\left\langle \partial_s\widehat{\Psi},\widehat{\Psi}\right\rangle.
$$
But
$$
\widehat{\Psi}(s,t)=\left|\widehat{\Psi}(s,t)\right| \exp\left(i\theta(s,t)\right),
$$
where
$$
\theta(s,t)=-\frac{1}{\hbar}Re(C(s))-\frac{\hbar}{2m}s^2t-\frac{1}{\hbar}\int V(t)dt.
$$
Hence 
$$
\partial_s\widehat{\Psi}(s,t)=\frac{Im(C'(s))}{\hbar}e^{Im(C(s))/\hbar}e^{i\theta(s,t)}+e^{Im(C(s))/\hbar}e^{i\theta(s,t)}i\partial_s\theta(s,t)\Rightarrow
$$
$$
\partial_{s}\widehat{\Psi}=e^{Im(C(s))/\hbar}e^{i\theta(s,t)}\left(\frac{Im(C'(s))}{\hbar}+i\partial_s\theta(s,t)\right)
$$
Hence
$$
\left|\partial_s\widehat{\Psi}\right|^2=e^{2Im(C(s))/\hbar}\left(\left(\frac{Im(C'(s))}{\hbar}\right)^2+\left(\frac{Re(C'(s))}{\hbar}+\frac{\hbar s t}{m}\right)^2\right).\eqno{(9)}
$$
Also
$$
(\partial_s\widehat{\Psi})\overline{\widehat{\Psi}}=\exp\left(\frac{2Im(C(s))}{\hbar}\right)\left(\frac{Im(C'(s))}{\hbar}-i\left(\frac{Re(C'(s))}{\hbar}+\frac{\hbar st}{m}\right)\right)=
$$
$$
=-i|\widehat{\Psi}|^2\frac{C'(s)}{\hbar}-\frac{i\hbar}{m}st|\widehat{\Psi}|^2=-i|\widehat{\Psi}|^2\left(\frac{C'(s)}{\hbar}+\frac{\hbar s t}{m}\right).\eqno{(10)}
$$
Hence we have
$$
\left\langle \textbf{x}\right\rangle=\frac{1}{2\pi\hbar}\int^{+\infty}_{-\infty}|\widehat{\Psi}|^2C'(s)ds+\frac{\hbar t}{2\pi m}\int^{+\infty}_{-\infty}s|\widehat{\Psi}|^2ds=
$$
$$
=\frac{1}{2\pi} \int^{+\infty}_{-\infty}|\widehat{\Psi}|^2\left(\frac{C'(s)}{\hbar}+\frac{\hbar st}{m}\right)ds=C_1+\frac{\hbar t}{2\pi m}\int^{+\infty}_{-\infty}s|\widehat{\Psi}|^2 ds
$$
and
$$
\frac{d\left\langle \textbf{x}\right\rangle}{dt}=\frac{\hbar}{2\pi m}\int^{+\infty}_{-\infty}s|\widehat{\Psi}|^2ds=constant.
$$
Hence the mean value of the position of the particle is lineary depended with time. If happens 
$$
\int^{+\infty}_{-\infty}s|\widehat{\Psi}|^2ds= \int^{+\infty}_{-\infty}s\cdot\exp\left(\frac{2Im(C(s))}{\hbar}\right)ds=0,
$$
then the position of particle is conserved. However the particle may move. This phenomenon is related with the parity operator. Since the Hamiltonian is independed on the change $x\rightarrow -x$, from (5), we have
$$
\Psi(-x,t)=\frac{1}{2\pi}\exp\left(-\frac{i}{\hbar}\int V(t)\right)\int_{\textbf{\scriptsize R\normalsize}}\exp\left(-\frac{i}{\hbar}C(s)-\frac{i\hbar}{2m}s^2t\right)e^{-isx}ds=
$$
$$
=\frac{1}{2\pi}\exp\left(-\frac{i}{\hbar}\int V(t)\right)\int_{\textbf{\scriptsize R\normalsize}}\exp\left(-\frac{i}{\hbar}C(-s)-\frac{i\hbar}{2m}s^2t\right)e^{isx}ds.
$$
Hence $C(-s)=C(s)$ if and only if $\Psi(-x,t)=\Psi(x,t)$. But then
$$
\int^{+\infty}_{-\infty}s|\widehat{\Psi}|^2ds=\int^{+\infty}_{-\infty}s\exp\left(2Im(C(s))/\hbar\right)ds=0
$$ 
and thus the position in this case is conserved.\\
Also
$$
\left\langle \textbf{x}^2\right\rangle=\left\langle x^2\Psi,\Psi\right\rangle=\frac{1}{2\pi}\left\langle \left(\widehat{x^2\Psi}\right),\widehat{\Psi}\right\rangle=\frac{1}{2\pi} \left\langle \int^{+\infty}_{-\infty}x^2\Psi(x,t)e^{-i x s}dx,\widehat{\Psi}\right\rangle=
$$
$$
=-\frac{1}{2\pi}\left\langle \partial^2_{ss}\widehat{\Psi},\widehat{\Psi}\right\rangle 
$$
and from (10) we have
$$
\left(\partial^2_{ss}\widehat{\Psi}\right)\overline{\widehat{\Psi}}+|\partial_s\widehat{\Psi}|^2=-\frac{2i}{\hbar^2}|\widehat{\Psi}|^2 Im(C'(s))C'(s)-\frac{i}{\hbar}|\widehat{\Psi}|^2C''(s)-
$$
$$
-\frac{i\hbar t}{m}|\widehat{\Psi}|^2-\frac{2ist}{m}|\widehat{\Psi}|^2Im(C'(s)).
$$
Hence using (9)
$$
\left(\partial^2_{ss}\widehat{\Psi}\right)\overline{\widehat{\Psi}}=-\frac{2i}{\hbar^2}|\widehat{\Psi}|^2 Im(C'(s))C'(s)-\frac{i}{\hbar}|\widehat{\Psi}|^2C''(s)-\frac{i\hbar t}{m}|\widehat{\Psi}|^2-
$$
$$
-\frac{2ist}{m}|\widehat{\Psi}|^2Im(C'(s))-|\widehat{\Psi}|^2\left(\frac{Im(C'(s))}{\hbar}\right)^2-|\widehat{\Psi}|^2\left(\frac{Re(C'(s))}{\hbar}\right)^2-
$$
$$
-|\widehat{\Psi}|^2\frac{\hbar^2s^2t^2}{m^2}-|\widehat{\Psi}|^2\frac{2Re(C'(s))st}{m}=
$$
$$
=-\frac{2st}{m}|\widehat{\Psi}|^2C'(s)+\frac{2}{\hbar^2}|\widehat{\Psi}|^2Im(C'(s))^2-\frac{2i}{\hbar^2}|\widehat{\Psi}|^2Im(C'(s))Re(C'(s))-
$$
$$
-\frac{i}{\hbar}|\widehat{\Psi}|^2C''(s)-\frac{i\hbar t}{m}|\widehat{\Psi}|^2-|\widehat{\Psi}|^2\frac{|C'(s)|^2}{\hbar^2}-|\widehat{\Psi}|^2\frac{\hbar^2s^2t^2}{m^2}=
$$
$$
=-\frac{2st}{m}|\widehat{\Psi}|^2C'(s)+\frac{1}{\hbar^2}|\widehat{\Psi}|^2Im(C'(s))^2+\frac{|\widehat{\Psi}|^2}{\hbar^2}Re(C'(s))^2-\frac{|\widehat{\Psi}|^2}{\hbar^2}C'(x)^2-
$$
$$
-\frac{i}{\hbar}|\widehat{\Psi}|^2C''(s)-\frac{i\hbar t}{m}|\widehat{\Psi}|^2-|\widehat{\Psi}|^2\frac{|C'(s)|^2}{\hbar^2}-|\widehat{\Psi}|^2\frac{\hbar^2s^2t^2}{m^2}=
$$
$$
=-\frac{2st}{m}|\widehat{\Psi}|^2C'(s)-\frac{|\widehat{\Psi}|^2}{\hbar^2}C'(s)^2-\frac{i|\widehat{\Psi}|^2}{\hbar}C''(s)-\frac{i\hbar t}{m}|\widehat{\Psi}|^2-|\widehat{\Psi}|^2\frac{\hbar^2s^2t^2}{m^2}=
$$
$$
=-|\widehat{\Psi}|^2\left(\frac{C'(s)}{\hbar}+\frac{st\hbar}{m}\right)^2-i|\widehat{\Psi}|^2\left(\frac{C''(s)}{\hbar}+\frac{\hbar t}{m}\right).\eqno{(11)}
$$
Hence
$$
\left\langle \textbf{x}^2\right\rangle=-\frac{1}{2\pi}\left\langle \partial^2_{ss}\widehat{\Psi},\widehat{\Psi}\right\rangle=\frac{1}{2\pi}\int^{+\infty}_{-\infty}|\widehat{\Psi}|^2\left(\frac{C'(s)}{\hbar}+\frac{st\hbar}{m}\right)^2ds+
$$
$$
+\frac{i}{2\pi}\int^{+\infty}_{-\infty}|\widehat{\Psi}|^2\left(\frac{C''(s)}{\hbar}+\frac{\hbar t}{m}\right)ds.
$$
Also
$$
i\hbar\frac{d}{dt}\left\langle \textbf{x}^2\right\rangle=\frac{i\hbar}{m\pi}\int^{+\infty}_{-\infty}C'(s)s|\widehat{\Psi}|^2ds-\frac{\hbar^2}{2\pi m}\int^{+\infty}_{-\infty}|\widehat{\Psi}|^2ds+\frac{it\hbar^3}{\pi m^2}\int^{+\infty}_{-\infty}|\widehat{\Psi}|^2s^2ds.\eqno{(11.1)}
$$
Finaly
$$
(\Delta \textbf{x})^2=\left\langle \textbf{x}^2\right\rangle-\left\langle \textbf{x}\right\rangle^2=\frac{1}{2\pi}\int^{+\infty}_{-\infty}|\widehat{\Psi}|^2\left(\frac{C'(s)}{\hbar}+\frac{st\hbar}{m}\right)^2ds+
$$
$$
+\frac{i}{2\pi}\int^{+\infty}_{-\infty}|\widehat{\Psi}|^2\left(\frac{C''(s)}{\hbar}+\frac{\hbar t}{m}\right)ds-\left(\frac{1}{2\pi}\int^{+\infty}_{-\infty}|\widehat{\Psi}|^2\left(\frac{C'(s)}{\hbar}+\frac{\hbar t s}{m}\right)ds\right)^2.\eqno{(12)}
$$
We have
$$
[\textbf{x}^2,\textbf{H}]\Psi=[\textbf{x x},\textbf{H}]\Psi=[\textbf{x},\textbf{H}] \textbf{x}\Psi+\textbf{x}[\textbf{x},\textbf{H}]\Psi=i\hbar\frac{\textbf{p}}{m}\textbf{x}\Psi+i\hbar\textbf{x}\frac{\textbf{p}}{m}\Psi=
$$
$$
=\frac{i\hbar}{m}\left(-i\hbar \partial_x(x\Psi)-i\hbar x\partial_x\Psi\right)=\frac{\hbar^2}{m}(\Psi+x\partial_x\Psi+x\partial_x\Psi)=
$$
$$
=\frac{\hbar^2}{m}(\Psi+2x\partial_x\Psi).
$$
Hence
$$
\left\langle [\textbf{x}^2,\textbf{H}] \right\rangle\Psi=\frac{\hbar^2}{m}(||\Psi||_2^2+2\left\langle x\partial_x\Psi,\Psi\right\rangle)=
$$
$$
=\frac{\hbar^2}{2\pi m}||\widehat{\Psi}||^2_2+\frac{2\hbar^2}{m}\left\langle x\partial_x\Psi,\Psi\right\rangle=\frac{\hbar^2}{2\pi m}\int^{+\infty}_{-\infty}|\widehat{\Psi}|^2ds
+\frac{\hbar^2}{\pi m}\left\langle\widehat{\partial_x \Psi},\widehat{x\Psi}\right\rangle=
$$
$$
=c_1+\frac{\hbar^2}{\pi m}\left\langle -(-is)\int^{+\infty}_{-\infty}\Psi(x,t)e^{-ixs}dx,i\int^{+\infty}_{-\infty}(-ix)\Psi(x,t)e^{-ixs}dx\right\rangle=
$$
$$
=c_1+\frac{\hbar^2}{\pi m}\left\langle s\widehat{\Psi},\partial_s\widehat{\Psi}\right\rangle=
$$
$$
=c_1+\frac{\hbar^2}{\pi m}\left\langle s |\widehat{\Psi}|e^{i\theta(s,t)},\partial_s(|\widehat{\Psi}|) e^{i\theta(s,t)}+|\widehat{\Psi}|e^{i\theta(s,t)}i\partial_s\theta(s,t)\right\rangle=
$$
$$
c_1+\frac{\hbar^2}{\pi m}\int^{+\infty}_{-\infty} s|\widehat{\Psi}|^2\frac{Im(C'(s))}{\hbar}ds-\frac{i\hbar^2}{\pi m}\int^{+\infty}_{-\infty} s |\widehat{\Psi}|^2\left(-\frac{1}{\hbar}Re(C'(s))-\frac{\hbar s t}{m}\right)ds=
$$
$$
=c_1+\frac{\hbar}{\pi m}\int^{+\infty}_{-\infty}s|\widehat{\Psi}|^2Im(C'(s))ds+\frac{i\hbar}{\pi m}\int^{+\infty}_{-\infty}s|\widehat{\Psi}|^2Re(C'(s))ds+
$$
$$
+\frac{i\hbar^3t}{\pi m^2}\int^{+\infty}_{-\infty}s^2|\widehat{\Psi}|^2ds.
$$
Hence
$$
\left\langle [\textbf{x}^2,\textbf{H}] \right\rangle\Psi=\frac{\hbar^2}{2\pi m}\int^{+\infty}_{-\infty}|\widehat{\Psi}|^2ds+\frac{i\hbar}{\pi m}\int^{+\infty}_{-\infty}s|\widehat{\Psi}|^2\overline{C'(s)}ds
+\frac{i\hbar^3t}{\pi m^2}\int^{+\infty}_{-\infty}s^2|\widehat{\Psi}|^2ds.\eqno{(13)}
$$
But there holds the general law
$$
i\hbar\frac{d}{dt}\left\langle \textbf{A}\right\rangle=\left\langle [\textbf{A},\textbf{H}]\right\rangle+i\hbar\left\langle \frac{\partial \textbf{A}}{\partial t}\right\rangle.\eqno{(14)}
$$
Now we now that
$$
\left\langle \textbf{x}^n\right\rangle=\int^{+\infty}_{-\infty}x^n|\Psi(x,t)|^2dx\in\textbf{R}.
$$
Also $\left\langle\frac{\partial \textbf{x}^n}{\partial t}\right\rangle=0$. Hence (14) is equivalent to
$$
i\hbar\frac{d}{dt}\left\langle\textbf{x}^2\right\rangle+\overline{\left\langle[\textbf{x}^2,\textbf{H}]\right\rangle}=0\eqno{(15)}
$$
Using (11.1) and (13) we can easily verify (15) and hence (14). Note here that we did not use (14) in first place because the function $C(s)$ satisfies conditions imposed by the Schrodinger equation and the nature of the problem and everytime we solve (1) we must find these conditions. For example $\left\langle \textbf{x}^n\right\rangle\in\textbf{R}\ldots \textrm{etc}$ gives some conditions on $C(s)\ldots$ etc.\\
Examples of (14) are:\\
\textbf{1)} Setting $\textbf{A}=\textbf{x}$, then 
$$
[\textbf{x},\textbf{H}]=i\hbar\frac{\partial \textbf{H}}{\partial p}=\frac{i\hbar}{m}\textbf{p}
$$
Hence
$$
i\hbar\frac{d\left\langle \textbf{x}\right\rangle}{dt}=\frac{i\hbar}{m}\left\langle\textbf{p}\right\rangle.
$$
Hence the position is not in general preserved.\\
\textbf{2)} Setting $\textbf{A}=\textbf{H}$, then (since $[\textbf{H},\textbf{H}]=0$):
$$
\frac{d\left\langle \textbf{E}\right\rangle}{dt}=\frac{d\left\langle \textbf{H}\right\rangle}{dt}=\left\langle\frac{\partial \textbf{H}}{\partial t}\right\rangle=\left\langle \frac{d\textbf{V}}{dt}\right\rangle.\textrm{ Energy not preserved. }
$$
\textbf{3)} If $\textbf{A}=\textbf{p}$, then from 
$$
[\textbf{p},\textbf{H}]=-i\hbar \frac{\partial \textbf{H}}{\partial x}=0,
$$ 
we have
$$
i\hbar\frac{d\left\langle \textbf{p}\right\rangle}{dt}=\left\langle [\textbf{p},\textbf{H}]\right\rangle+i\hbar\left\langle \frac{\partial \textbf{p}}{\partial t}\right\rangle=0.\textrm{ Momentum is always preserved. }
$$
\textbf{4)} If $\textbf{P}=$parity operator, then 
$$
\left\langle [\textbf{P},\textbf{H}]\right\rangle=\textbf{P}\textbf{H}\Psi(x,t)-\textbf{H}\textbf{P}\Psi(x,t)=\textbf{P}\textbf{E}\Psi(x,t)-\textbf{H}\Psi(-x,t)-
$$
$$
=i\hbar\partial_t\Psi(-x,t)-\textbf{H}\Psi(-x,t)=i\hbar\partial_t\Psi_e(-x,t)+i\hbar\partial\Psi_o(-x,t)+
$$
$$
+\frac{\hbar^2}{2m}\partial^2_{xx}\Psi_e(-x,t)+\frac{\hbar^2}{2m}\partial^2_{xx}\Psi_o(-x,t)
-V(t)(\Psi_e(-x,t)+\Psi_o(-x,t))
$$ 
The above quantity is zero in either case of $\Psi=\Psi_e$ (even) and $\Psi=\Psi_o$ (odd). Hence the parity is always preserved.

\[
\]

\centerline{\bf References}

[1]: Nirmala Prakash. ''Differential Geometry An Integrated Approach''. Tata McGraw-Hill Publishing Company Limited. New Delhi. 1981.

[2]: Murray R. Spiegel. 'Schaum's Outline of Theory and Problems of Fourier Analysis with Applications to Boundary Value Problems'. McGraw-Hill, New York. 1974.

[3]: Frigyes Riesz and Bela Sz.-Nagy. 'Functional Analysis'. Dover Publications, Inc. New York. 1990. 

[4]: Eduard Prugovecki. 'Quantum Mechanics in Hilbert Space'. Academic Press. New York and London. 1971.

[5]: Jerome A. Goldstein. 'Semigroups of Linear Operators and Applications'. Oxford University Press-New York. Clarendon Press-Oxford. 1985.

\end{document}